\documentclass[11pt]{article}

\usepackage[latin1]{inputenc}

\usepackage{amstext,amsmath,amssymb,amsfonts,bbm,amsmath}
\usepackage{dsfont}
\usepackage{extarrows}
\usepackage{xcolor} 
\usepackage{hyperref}
\hypersetup{
    colorlinks=false,
    menubordercolor=red,
    linkbordercolor=blue
}
\usepackage{authblk}
\usepackage{caption} 
\usepackage{epsfig} 
\usepackage{color} 
\usepackage{graphicx} 
\usepackage{braket} 
\usepackage{slashed} 
\usepackage{dsfont} 
\usepackage{amsthm}  
\usepackage{multirow}
\usepackage{longtable}
\usepackage{mathtools}

\usepackage{cite}

\usepackage{psfrag}
\usepackage{tikz} 
\usetikzlibrary{arrows}
\usetikzlibrary{plotmarks}
\usetikzlibrary{external}
\usetikzlibrary{patterns}
\usepackage{pgfplots}
\pgfplotsset{compat=1.9}
\usetikzlibrary{patterns}
\usetikzlibrary{calc}

\usepackage{float}  
\usepackage{subfig}
\usepackage[lmargin=60pt,rmargin=60pt,tmargin=60pt,bmargin=65pt]{geometry}
\usetikzlibrary{calc, external, intersections, arrows.meta, arrows, patterns, plotmarks, shapes.geometric, pgfplots.groupplots}

\pgfplotsset{width=10cm, compat=1.9,every non boxed y axis/.style={}}
\usepgfplotslibrary{colorbrewer}

\hypersetup{
           pdftitle={Novel Defect Universality Classes from Interacting RG Interfaces},
           pdfdisplaydoctitle
}

\tikzset{cross/.style={path picture={
      \draw[black]
            (path picture bounding box.south east) --
            (path picture bounding box.north west)
            (path picture bounding box.south west) --
            (path picture bounding box.north east);}}}

\newcommand{\lsp}{\hspace{0.5pt}}

\allowdisplaybreaks[4]

\numberwithin{equation}{section}

\theoremstyle{remark}

\def\pe{\phantom{{}={}}}

\DeclareMathAlphabet{\mathdutchcal}{U}{dutchcal}{m}{n}
\SetMathAlphabet{\mathdutchcal}{bold}{U}{dutchcal}{b}{n}
\DeclareMathAlphabet{\mathdutchbcal}{U}{dutchcal}{b}{n}

\begin{document}

\title{\bf Novel Defect Universality Classes from Interacting RG Interfaces }

\author[1]{Samuel Bartlett-Tisdall}
\author[2]{Sabine Harribey}
\author[1]{William H.\ Pannell}

\affil[1]{\normalsize\it 
Department of Mathematics, King's College London, Strand, London WC2R 2LS, United Kingdom
\authorcr \hfill }

\affil[2]{\normalsize \it 
School of Theoretical Physics, Dublin Institute For Advanced Studies, 10 Burlington Road, Dublin, Ireland
  \authorcr \hfill}

\date{}

\maketitle

\hrule\bigskip

\begin{abstract}
We search for new defect universality classes by considering localised interactions placed on an RG interface separating two interacting multiscalar CFTs in $4-\varepsilon$ dimensions. Studying interactions spread throughout the entire interface as well as defects restricted to lines and surfaces within the interface, we find that this setup leads to a great number of additional physical fixed points in the space of conformal defects. At one loop it is possible to interpret these fixed points as coming from defects placed within a single bulk whose interaction is an average of the two sides. This averaging means that it is possible to identify conformal defects with considerably less global symmetry than was possible beforehand. We finally compute conformal data for this setup, and find the free energy associated with these RG interfaces.
\end{abstract}

\hrule\bigskip

\tableofcontents

\section{Introduction}
\label{sec:introduction}




Introducing defects in quantum field theories (QFTs) is useful to describe numerous physical systems such as impurities or localised interactions\footnote{For experimental realisations of such systems, see for example \cite{sigl1986order, mailander1990near, burandt1993near, alvarado1982surface, PhysRevA.19.866, PhysRevB.40.4696, PhysRevB.58.12038}} and can also be studied from a purely theoretical standpoint to explore new regimes of QFTs, often analytically accessible. 
When applied to conformal field theories (CFTs), such defects give rise to new types of CFTs called defect conformal field theories (dCFTs), realising conformal invariance in a smaller dimension, thus expanding the landscape of  universality classes \cite{McAvity:1995zd,Billo:2016cpy}. The resulting dCFTs have been studied extensively in recent years. In the case of the $O(N)$ model, research dates back many years \cite{AJBray_1977, Ohno:1983lma, gompper1985conformal, McAvity:1995zd, Diehl:1996kd} and recent renewed interest has uncovered new universality classes \cite{Metlitski:2020cqy, Padayasi:2021sik, Toldin:2021kun, Krishnan:2023cff,DeSabbata:2025ano,Giombi:2023dqs, Trepanier:2023tvb, Raviv-Moshe:2023yvq,Harribey:2023xyv}. More recently, defects with a continuously adjustable dimension $p=2+\delta$ were considered in \cite{deSabbata:2024xwn}, while line and surface defects for the long-range $O(N)$ model were considered in \cite{Bianchi:2024eqm}.

Going beyond the $O(N)$ model, the epsilon expansion can be used to study a wide range of CFTs that are reached as endpoints of renormalisation group (RG) flows triggered by operators that break the $O(N)$ symmetry. RG fixed points for various global symmetries can then be investigated by generalising to quartic multiscalar models \cite{Pelissetto:2000ek, Osborn:2017ucf, Rychkov:2018vya, Osborn:2020cnf}. 
RG flows and CFTs that emerge when these theories are deformed by defects have been studied extensively, for instance for line defects in  \cite{Pannell:2023pwz}, for surface defects in \cite{Anataichuk:2025zoq}, and for interfaces in \cite{Harribey:2024gjn}.

The global symmetry breaking patterns induced by the introduction of a defect is highly dependent on the starting global symmetry of the bulk, and generally increases in complexity as one moves from a line defect to a surface defect to an interface. In an $O(N)$ bulk, for instance, a line defect will simply pin the order parameter along a certain vector, and can thus only break the symmetry to $O(N-1)$, while a surface can break the symmetry to $O(p)\times O(N-p)$ for any $p\leq N$, and an interface can break the symmetry to a discrete subgroup. For more complicated bulk symmetries, such as a hypertetrahedral bulk, more patterns of symmetry breaking emerge\cite{Pannell:2024hbu}. Careful analysis of the global symmetries which these defects can have is not of just theoretical interest, and also has experimental applications. Indeed, in three dimensions the $O(3)$ and Cubic universality classes are practically indistinguishable as their critical exponents are very close. Introducing a line defect will break the bulk symmetries in different ways and thus allow to distinguish the two universality classes without measuring critical exponents.

Nevertheless, the types of symmetry groups one can begin with in the bulk are considerably limited for multiscalar models\cite{Osborn:2020cnf}, and it thus appears that the global symmetries present in conformal scalar defects is similarly limited. We wish to present a simple model which goes beyond this paradigm, and demonstrate that it is possible to construct conformal defects with considerably less remaining global symmetry than before, and even in some cases with no global symmetry at all. In this paper, we will consider the setup where there is not just a single bulk CFT, but rather two, which are separated by a so-called RG interface. This can be described by the bulk action
\begin{equation}
    S=\int d^d x \left( \frac{1}{2}\partial^{\mu} \phi_i\partial_{\mu} \phi_i + \theta (x_{\perp})\frac{\lambda^1_{ijkl}}{4!} \phi_i\phi_j\phi_k\phi_l + \theta (-x_{\perp})\frac{\lambda^2_{ijkl}}{4!} \phi_i\phi_j\phi_k\phi_l  \right) \, .
\label{eq:actionbulk}
\end{equation}
The interface is located at $x_{\perp}=0$, for $x_{\perp}>0$ we have $\text{CFT}_1$ with coupling $\lambda^1_{ijkl}$ and for $x_{\perp}<0$ we have $\text{CFT}_2$ with coupling $\lambda^2_{ijkl}$. When one of the CFTs is a free theory, the cases of the $O(N)$ model and of double trace deformations have been considered in \cite{Gliozzi:2015qsa,Giombi:2024qbm}. RG interfaces have also been studied much more extensively in two dimensions, see for example \cite{Gaiotto:2012np,Quella:2006de,Cogburn:2023xzw,Konechny:2020jym}, or in the context of supergravity, see for example \cite{HultgreenMena:2025uok} and references therein.

We will then add interactions on the interface and compute fixed points for the interface couplings as well as CFT data. We will consider three types of interactions on the interface: cubic couplings on the interface, line  sub-defects, and surface sub-defects, given respectively by
\begin{equation}\label{eq:dacti}
    S_{\text{interface}}=\int d^{d-1}x \frac{h_{ijk}}{3!}\phi_i\phi_j\phi_k \, ,
\end{equation}
\begin{equation}\label{eq:dactl}
    S_{\text{line}}= \int dx_1 \,  h_i \phi_i \, ,
\end{equation}
and
\begin{equation}\label{eq:dacts}
    S_{\text{surface}}= \int dx_1 dx_2\frac{h_{ij}}{2}\phi_i\phi_j \, .
\end{equation}
If we denote $x=(x_{\parallel},x_{\perp})$ and $x_{\parallel}=(x_1,x_2,\vec{x})$, the line sub-defect is located at $x_{\perp}=0 \, , x_2=0 \, ,\vec{x}=0 $ and the surface sub-defect is located at $x_{\perp}=0 \,  ,\vec{x}=0 $.

We could study these models with the folding trick as in \cite{Giombi:2024qbm,Gliozzi:2015qsa}. However, because we have interactions located on the interface, we would have to be careful about boundary conditions and distinction between the original right and left fields. Without the folding trick, we do not have to worry about boundary conditions, but we have different propagators: bulk-to-bulk, bulk-to-interface, and interface-to-interface, as well as different vertices on each side of the interface. In the rest of the paper, we will do all computations perturbatively at leading order, following the method of \cite{Gliozzi:2015qsa}.

The rest of the paper is organised as follows.  In section \ref{sec:betafunctions}, we compute beta functions and fixed points for each model: cubic interface, line sub-defect, and surface sub-defect. We discuss the additional fixed points that appear in these mixed bulk theories. In section \ref{sec:data}, we compute CFT data for the cubic interface: one-point functions, two-point functions, and free energy. We end with a discussion in section \ref{sec:discussion}, and comment on future directions. We consider a variety of non-trivial classes of quartic interactions, which we briefly summarise in appendix \ref{sec:appendix}.

\section{Beta Function}
\label{sec:betafunctions}
In this section, we are interested in finding critical dCFTs in our model (\ref{eq:actionbulk}) in $d=4-\varepsilon$ dimensions. In general, one can consider a bulk action $S(\lambda^I)$, that depends on some couplings $\lambda^I$. If $\lambda^I$ couples to some nearly-marginal operators, then these bulk couplings will undergo an RG flow to an IR fixed point $(\lambda^*)^I$. At such a point, we have a CFT $S_{CFT}((\lambda^*)^I)$. In our work, we consider multiscalar theories defined in the bulk by
\begin{equation}
    S=\int d^d x \frac{1}{2}\partial^{\mu} \phi_i\partial_{\mu} \phi_i + \frac{\lambda_{ijkl}}{4!} \phi_i\phi_j\phi_k\phi_l\,.
\end{equation}
The RG flow is governed by the beta function, a differential equation relating the couplings to a characteristic energy scale $\mu$, which at the one-loop level is
\begin{equation}
    \beta_{ijkl}(\lambda) = \frac{d \lambda_{ijkl}}{d \ln\mu} =-\varepsilon\lambda_{ijkl}+(\lambda_{ijmn}\lambda_{mnkl}+\text{Perms.}) \, .
\end{equation}
Solving for fixed points of the beta function $\beta_{ijkl} = 0$ provides critical couplings $\lambda^*_{ijkl}$ where the bulk theory is a CFT.

Consider that we have fixed a bulk CFT. To construct a defect within the bulk theory, we integrate operators $\mathcal{O}_I$ over a submanifold $\mathcal{M}$,
\begin{equation}
    S_{\text{defect}}=\int_\mathcal{M} h^I\mathcal{O}_I\,,
\end{equation}
with defect couplings $h^I$. If these defect operators are nearly-marginal, then the defect couplings $h^I$ will undergo their own defect RG flow to IR fixed points, in which case we obtain a conformal defect, a dCFT. As in the bulk case, the flow is governed by its own defect beta function
\begin{equation}
    \frac{dh^I}{d\ln\mu}=\beta^I(h,\lambda^*)\,.
\end{equation}
Note that this flow will depend on the bulk CFT, defined by the critical coupling $\lambda^*$. Such a dCFT is then defined by a critical coupling $(h^*)^I(\lambda^*)$.

Alternatively, we can forget for the moment that the bulk is critical. Then solving for critical points of the defect beta function gives us continuous families $(h^*)^I(\lambda)$, depending on the (generally non-critical) bulk coupling $\lambda$. This is a much larger space of solutions for $(h^*)^I$. From our discussion above, we know that not all of these defect couplings will be critical. However, the result we show in this section is that there are additional points in this solution space that occur for some $\lambda \neq \lambda^*$. With the interface theory (\ref{eq:actionbulk}), specifically we find that the defect beta function at one loop only depends upon the bulk couplings through their average $\lambda_{ijkl} = \frac{\lambda^1_{ijkl}+\lambda^2_{ijkl}}{2}$ (note that $\lambda^1$ and $\lambda^2$ are themselves critical). This novel physical picture adds a great number of fixed points to the space of conformal defects. This observation is only a one-loop trick, as at higher orders it may no longer be possible to obtain the defect beta function within the RG interface from the usual defect beta function with the assignment $\lambda_{ijkl} = \frac{\lambda^1_{ijkl}+\lambda^2_{ijkl}}{2}$. However, higher-loop corrections to the defect coupling will be completely determined in terms of the one-loop results, and thus this picture will still be powerful when including higher-loop effects. It is worthwhile to note that it is possible to distinguish between defect fixed points in a bulk CFT and in an RG interface by the fact that the symmetry breaking of the conformal symmetry will be different. In the former case, a $p$-dimensional defect will induce the breaking
\begin{equation}
    SO(d+1,1)\rightarrow SO(p+1,1)\times SO(d-p)\,,
\end{equation}
while in the latter case the symmetry will be broken like
\begin{equation}
    SO(d+1,1)\xrightarrow{\text{RG Interface}} SO(d,1)\xrightarrow{\text{Conformal Defect}} SO(p+1,1)\times SO(d-1-p)\,.
\end{equation}

As we will often be working with general values for $\lambda_{ijkl}$, we will make a distinction between couplings satisfying $\beta(h^*,\lambda)=0$ for generic $\lambda_{ijkl}$, and couplings satisfying $\beta(h^*,\lambda)=0$ for special $\lambda$ which can be given a physical interpretation, either in terms of a conformal defect in a single bulk, in which case $\lambda^*$ is a fixed point of the bulk beta function (\ref{eq:betabulk}), or a conformal composite defect within an RG interface, in which case  $\lambda_{ijkl} = \frac{\lambda^1_{ijkl}+\lambda^2_{ijkl}}{2}$ for conformal $\lambda^1$ and $\lambda^2$. The former we will call `solutions' or `families of solutions', and we will reserve the term `fixed point' or `physical' for defects in the second class.

\subsection{Cubic Interface}
The most natural choice of the submanifold $\mathcal{M}$ is that of the RG interface itself $\mathcal{M}=\{x\in\mathbb{R}^d\,|\,x_\perp=0\}$. As we are working in $d=4-\varepsilon$, the classically marginal operator will be cubic in the basic field $\phi_i$, leading to defect action given in \eqref{eq:dacti}. The model will differ only from the cubic interface considered in \cite{Harribey:2024gjn} in that the integrals over bulk points must be divided into two regions, $x_{\perp}>0$ where the quartic coupling is $\lambda^1_{ijkl}$, and then $x_{\perp}<0$ where the quartic coupling is $\lambda^2_{ijkl}$.

As in \cite{Harribey:2024gjn}, the introduction of cubic couplings on the interface will induce an RG flow, governed by a beta function $\beta_{ijk}=d h_{ijk}/d\ln{\mu}$ which can be determined by renormalizing the one-point function $\langle\phi_i\phi_j\phi_k(x)\rangle$. The purely defect diagram arising at one loop will not feel the presence of the two bulk couplings,
\begin{equation}
    \begin{tikzpicture}[baseline=(vert_cent.base)]
        \node at (90:0.6cm) [circle,draw,fill=white,inner sep=1.2pt,outer sep=0pt]  (t) {};
        \node at (210:0.6cm) [circle,draw,fill=white,inner sep=1.2pt,,outer sep=0pt] (l) {};
        \node at (330:0.6cm) [circle,draw,fill=white,inner sep=1.2pt,,outer sep=0pt] (r) {};
        \draw[very thick] (t)--(l);
        \draw[very thick] (l)--(r);
        \draw[very thick] (r)--(t);
        \node (vert_cent) at (current bounding box.center) {};
        \draw[very thick] (t)--++(90:0.4cm);
        \draw[very thick] (l)--++(210:0.4cm);
        \draw[very thick] (r)--++(330:0.4cm);
    \end{tikzpicture}
=\frac{h_{ilm}h_{jln}h_{kmn}}{16\pi^{2}\varepsilon} +\text{O}(\varepsilon^0)\,.
\end{equation}
The other one-loop diagram, involving a mixing of bulk and interface vertices, must be modified to
\begin{equation}
\begin{split}
    \begin{tikzpicture}[baseline=(vert_cent.base), square/.style={regular polygon,regular polygon sides=4}]
        \node at (0,0) [square,draw,fill=white,inner sep=1.2pt,outer sep=0pt]  (l) {};
        \node at (1.2,0) [circle,draw,fill=white,inner sep=1.2pt,outer sep=0pt] (r) {};
        \draw[very thick] (l) to [out=30,in=150] (r);
        \draw[very thick] (l) to [out=-30,in=-150] (r);
        \draw[very thick] (l)--++(150:0.4cm);
        \draw[very thick] (l)--++(-150:0.4cm);
        \draw[very thick] (r)--++(0:0.4cm);
        \node[inner sep=0pt,outer sep=0pt] (vert_cent) at (0,0) {$\phantom{\cdot}$};
        \node[xshift=-5pt] at (-0.4,0.4) {$i$};
        \node[xshift=-5pt] at (-0.4,-0.4) {$j$};
        \node[xshift=5pt] at (1.6,0) {$k$};
    \end{tikzpicture}=&\frac{\mu^{-\varepsilon}}{2}\lambda^1_{ijmn}h_{mnk}\int_0^\infty dy\int\frac{d^{d-1}\vec{p}}{(2\pi)^{d-1}}\frac{e^{-2(|\vec{p}|+|\vec{q}|)|y|}}{(2|\vec{p}|)^2}\\&+\frac{\mu^{-\varepsilon}}{2}\lambda^2_{ijmn}h_{mnk}\int_{-\infty}^0 dy\int\frac{d^{d-1}\vec{p}}{(2\pi)^{d-1}}\frac{e^{-2(|\vec{p}|+|\vec{q}|)|y|}}{(2|\vec{p}|)^2}\\
    =&\frac{1}{32\pi^2\varepsilon}(\lambda^1_{ijmn}h_{mnk}+\lambda^2_{ijmn}h_{mnk})\,.
\end{split}
\end{equation}
After rescaling the couplings, one thus finds the following modified beta function
\begin{align}\label{eq:betaone}
\beta_{ijk} &= -\frac{\varepsilon}{2} h_{ijk}
 +  \frac{1}{2}\big((\lambda^1_{ijmn}h_{mnk}+\lambda^2_{ijmn}h_{mnk})+ \textrm{Perms.} \big) -\frac{1}{4}h_{ilm}h_{jln}h_{kmn} \, .
\end{align}
This differs from the beta function considered in \cite{Harribey:2024gjn} only by the replacement $\lambda_{ijkl}\rightarrow\tfrac{\lambda^1_{ijkl}+\lambda^2_{ijkl}}{2}$, and naturally matches that beta function for the case of a single bulk $\lambda^1=\lambda^2$. One sees how considering the case of an RG interface dramatically increases the space of physical defect solutions, as we can make sense of the interface defect beta function for $\lambda^1\neq \lambda^2$, while before only the special case $\lambda^1= \lambda^2$ was considered. Now, as long as the tensor $\lambda_{ijkl}$ lies at the average of two bulk fixed points, the solution $h^*_{ijk}(\lambda)$ will be physically meaningful.

\subsubsection{Analytic Fixed Points}
One can ask what happens to the analytic fixed points identified in \cite{Harribey:2024gjn} for a single bulk, which is equivalent here to the case $\lambda^1=\lambda^2$. Those points were constructed by taking the interface interaction to lie in the rank-3 symmetric, traceless representation of some group, denoted $h_{ijk}=\mathdutchcal{h}\lsp d_{ijk}$, and assuming that the group has no rank-2 invariant tensor other than $\delta_{ij}$ so that there exist the relations
\begin{equation}\label{eq:analytichrelations}
    d_{ikl}d_{jkl}=r\lsp\delta_{ij}\,,\quad r>0\,,\qquad d_{ilm}d_{jln}d_{kmn}=s\lsp d_{ijk}\,,
\end{equation}
where the constants $r$ and $s$ depend upon the particular choice of group. To see if these fixed points survive in the RG interface setup, let us take $\lambda^1$ to lie at the $O(N)$ fixed point, (\ref{eq:lambdaON}), and $\lambda^2$ to be the Gaussian free theory, $\lambda^2_{ijkl}=0$. One can then use \eqref{eq:analytichrelations} to reduce the interface beta function \eqref{eq:betaone} to the simpler form
\begin{equation}
    \beta_{ijk}=\left(-\frac{N+2}{2(N+8)}\varepsilon\mathdutchcal{h}-\frac{1}{4}s\mathdutchcal{h}^3\right)d_{ijk}.
\end{equation}
This has the non-trivial, stable fixed point
\begin{equation}
    \mathdutchcal{h}^2=-\frac{2(N+2)}{s(N+8)}\varepsilon\,,
\end{equation}
which is unitary only for $s<0$. Examining the values of $r$ and $s$ given for various groups in \cite{Harribey:2024gjn}, one sees that for $N=8$ there will exist a unitary fixed point with $SU(3)$ symmetry. Thus, the fixed point survives unscathed, apart from small numerical differences.

\paragraph{Infinitesimal twisting:}
Bulk fixed points do not exist in isolation, but instead lie on an $O(N)$ orbit, the form of which depends upon the unbroken subgroup $H\subset O(N)$. This orbit is spanned by the action of the broken $O(N)$ symmetry generators, i.e. it is the set
\begin{equation}
    \{\lambda'_{ijkl}=R_{ia}R_{jb}R_{kc}R_{ld}\lambda_{abcd}|R\in O(N)\}\,.
\end{equation}
One can then ask what happens to the cubic interface when the different halves of space are twisted in relation to one another by one of these $O(N)$ field redefinitions. The simplest such case is the one in which the rotation connecting the two halves is infinitesimal, so that
\begin{equation}    \lambda^2_{ijkl}=\lambda^1_{ijkl}+\left(\omega_{ia}\lambda^1_{ajkl}+\text{Perms.}\right)+O(\omega^2)\,,
\end{equation}
where $\omega_{ij}$ is infinitesimal and antisymmetric. To understand this twisting let us first understand the effect $O(N)$ field redefinitions have on the usual interface setup. If the bulk was entirely $\lambda^2$, then one can see that the interface beta function \eqref{eq:betaone} will also rotate to
\begin{equation}
    \beta_{ijk}\rightarrow \beta_{ijk}+\left(\omega_{ia}\beta_{ajk}+\text{Perms.}\right)+O(\omega^2)
\end{equation}
so long as we also rotate the defect interaction tensor
\begin{equation}
    h_{ijk}\rightarrow h_{ijk}+\left(\omega_{ia}h_{ajk}+\text{Perms.}\right)+O(\omega^2)\,.
\end{equation}
We then see that rotating the bulk simply has the effect of rotating all of the fixed points, as one might expect.

Splitting the bulk back into $\lambda^1$ and $\lambda^2$, and then taking $\lambda^2$ to be an infinitesimal rotation away from $\lambda^1$, we see that from the perspective of the interface, the beta function \eqref{eq:betaone} is equivalent to taking the interface to be in a single half-rotated bulk,
\begin{equation}
    \lambda^3_{ijkl}=\frac{\lambda^1_{ijkl}+\lambda^2_{ijkl}}{2}=\lambda^1_{ijkl}+\left(\frac{\omega_{ia}}{2}\lambda_{ajkl}+\text{Perms.}\right)+O(\omega^2)\,.
\end{equation}
Thus, all cubic interface fixed points in the infinitesimally twisted bulk can be found at one loop by taking fixed points in a single $\lambda^1$ bulk and then performing half of that infinitesimal rotation on them.

\subsubsection{Numerical Fixed Points With Small \texorpdfstring{$N$}{N}}
\begin{figure}[t]
\centering
\begin{tikzpicture}
\begin{axis}[
    xmin=-0, xmax=6,
    ymin=0, ymax=6,
    xlabel= $h^2$,
    ylabel= $|h_i|^2$,
    ylabel style={rotate=-90},
    title={$O(2)$--Free Interface},
    legend pos=outer north east,
    reverse legend
]
\addplot+[
    no marks,
    color=gray,
    dotted,thick]
table{Data/02Fintp1.dat};
\addplot+[
    no marks,
    color=gray,
    dotted,thick]
table{Data/02Fintp2.dat};
\addplot[only marks,mark=square*,mark options={color=Dark2-A},mark size=2pt] coordinates {(0,0)};
\addlegendentry{Free bulk and $O(2)$--Free interface}
\addplot[only marks,mark=*,mark options={color=red},mark size=2pt] coordinates {(3.56128, 4.53274) (1.71629, 0.484939) (2.72243, 0.582321)};
\addlegendentry{$O(2)$ bulk}
\end{axis}
\end{tikzpicture}
    \caption{Fixed points of \eqref{eq:betaone}, with the choice of bulk tensor \eqref{eq:lambdachoiceOFint}. The dotted lines indicate the trajectories of solutions for arbitrary $\lambda$. The red circles indicate the physical fixed points in an $O(2)$ bulk, with $\lambda=1/10$, while the green square indicates the trivial defect solution, which is the only fixed point in both the Free bulk and $O(2)$--Free interface. Examining the trajectories, one sees that two of the $O(2)$ fixed points merge at $\lambda=1/12$ and move into the complex plane before they can produce a non-trivial fixed point in the $O(2)$--Free interface.}
\label{fig:OF2int}
\end{figure}
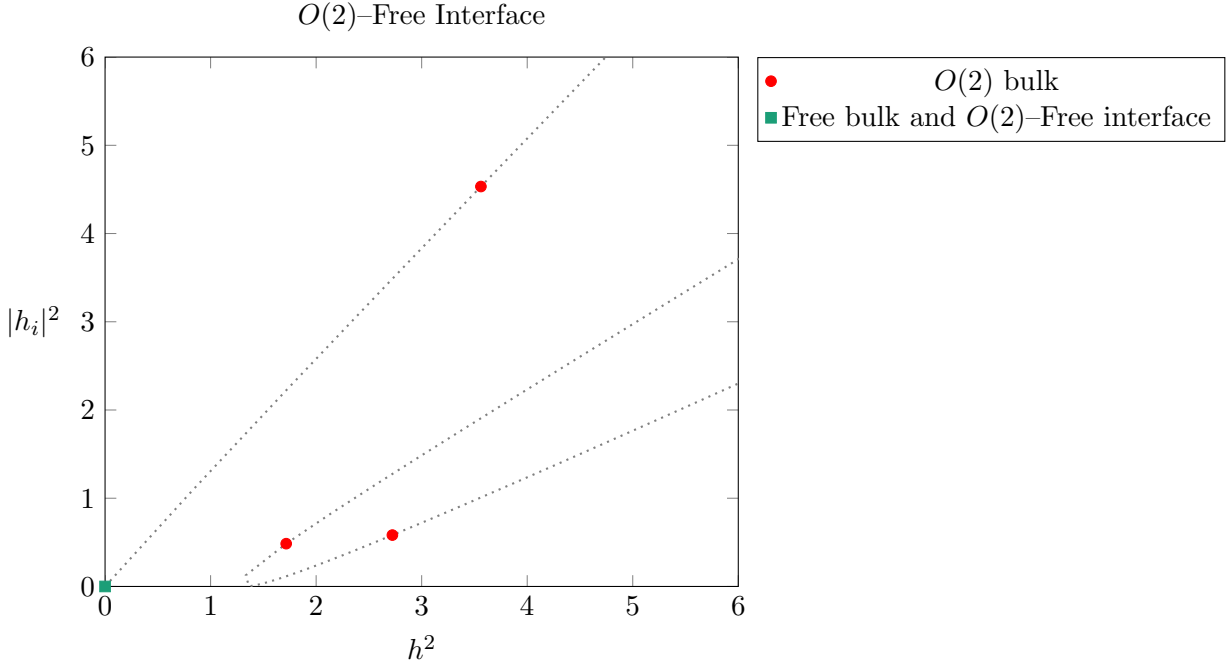
When there are only a few scalar fields in the system, the number of equations in the beta function \eqref{eq:betaone} is small enough to be tractable while one still finds enough fixed points to get a picture of what is going on. Let us first consider the simple case where one of the CFTs is taken to be free. For $N=1$ it is straightforward to see that there are no non-trivial fixed points. Taking $\lambda^1=0$, $\lambda^2=\varepsilon/3$,
the beta function becomes
\begin{equation}
    \beta_h=-\frac{\varepsilon}{2}h-\frac{1}{4}h^3+\frac{3}{2}\lambda h=-\frac{1}{4}h^3\,,
\end{equation}
which has no real fixed points beyond the trivial one, $h=0$. For $N=2$ we can consider a number of different interfaces: the $O(2)$--Free interface, the $O(2)$--$I\times I$ interface and the $I\times I$--Free interface, where $I\times I$ is the theory consisting of two decoupled Ising models. Looking at the beta function in each of these cases one finds again that there are no real non-trivial fixed points. The lack of non-trivial fixed points in these RG interfaces can be seen to be a consequence of the lack of non-trivial interfaces in a free theory for $N<3$. To illustrate this, we first examine the $O(2)$--Free interface more closely. Rather than solving for fixed points with a conformal bulk tensor, let us examine \eqref{eq:betaone} for
\begin{equation}\label{eq:lambdachoiceOFint}
    \lambda^1_{ijkl}=\lambda^2_{ijkl}=\varepsilon\lambda(\delta_{ij}\delta_{kl}+\text{Perms.})\,.
\end{equation}
As $\lambda$ changes from $1/10$, the $O(2)$ fixed point, it will pass through the $O(2)$--Free interface at $\lambda=1/20$, before reaching the free theory at $\lambda=0$. Solving numerically for various values of $\lambda$, one obtains the picture shown in Figure \ref{fig:OF2int}, where we have used the $O(N)$ invariants
\begin{equation}\label{eq:hinvariants}
h^2=h_{ijk}h_{ijk}\qquad\qquad|h_i|^2=h_{ijj}h_{ikk}\,.
\end{equation}
For $\lambda=1/10$ there are the three interface fixed points identified in \cite{Harribey:2024gjn}. As $\lambda$ decreases, one of these moves towards the origin, merging with the free defect at $\lambda=1/20$, while the other two trajectories merge at $\lambda=1/12$, and then move into the complex plane. This picture remains true when placing another non-free CFT on one side of the interface, with all of the non-trivial fixed points either merging to become complex, or becoming trivial themselves. The precise trajectories we have constructed pass through various critical points at the one-loop level. Such curves will exist for higher loops, though the form will be more complicated. This also applies to the subsequent trajectories we consider. 

\begin{figure}[t]
\centering
\begin{tikzpicture}
\begin{axis}[
    xmin=-0, xmax=6,
    ymin=0, ymax=6,
    xlabel= $h^2$,
    ylabel= $|h_i|^2$,
    ylabel style={rotate=-90},
    title={$O(2)$--$I\times I$ Interface},
    legend pos = outer north east,
]
\addplot[only marks,mark=*,mark options={color=red},mark size=2pt] coordinates {(3.56128, 4.53274) (1.71629, 0.484939) (2.72243, 0.582321)};
\addlegendentry{$O(2)$ bulk \cite{Harribey:2024gjn}};
\addplot[only marks,mark=triangle*,mark options={color=blue},mark size=2pt] coordinates {(3.69208, 4.24752) (4.30792, 0.552479) (2.56383, 0.559602) (2.43617, 3.2404)};
\addlegendentry{$O(2)$--$I\times I$ interface};
\addplot[only marks,mark=square*,mark options={color=Dark2-A},mark size=2pt] coordinates {(0,0) (2,2) (4,4)};
\addlegendentry{$I\times I$ bulk \cite{Harribey:2024gjn}};
\addplot+[
    no marks,
    color=gray,
    dotted,thick]
table{Data/O2Cintp1.dat};
\addplot+[
    no marks,
    color=gray,
    dotted,thick]
table{Data/O2Cintp2.dat};
\addplot+[
    no marks,
    color=gray,
    dotted,thick]
table{Data/O2Cintp3.dat};
\addplot+[
    no marks,
    color=gray,
    dotted,thick]
table{Data/O2Cintp4.dat};
\end{axis}
\end{tikzpicture}
    \caption{Fixed points of \eqref{eq:betaone} in the $O(2)$--$I\times I$ system. The dotted lines indicate the families of solutions parametrised by $\alpha$ with the choice of bulk tensor \eqref{eq:lambdachoiceOB2int}. One sees that the two non-trivial fixed points in the $I\times I$ bulk both lead, in the $\alpha\rightarrow 1$ limit, to one of the three $O(2)$ fixed points. the other two $O(2)$ fixed points instead have their origin in the region of strong coupling.}
\label{fig:OC2int}
\end{figure}
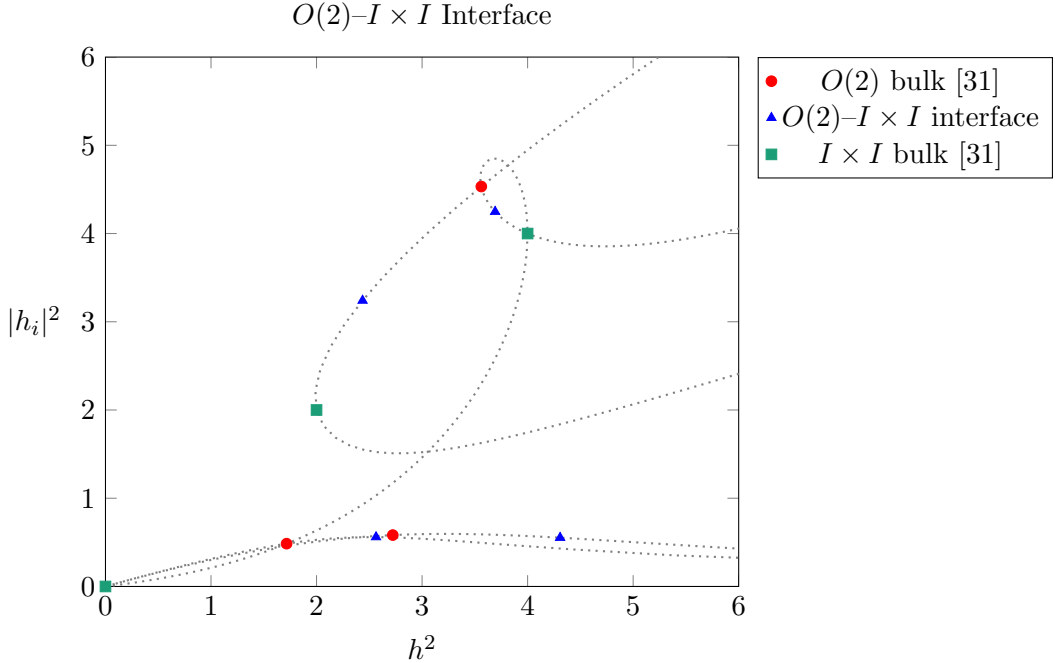

Secondly, we consider both of the CFTs to be interacting, in which case the existence of non-trivial fixed points in both bulks produces a richer set of trajectories. As an example, let us consider the case where one side lies in the $O(2)$ universality class, while the other side lies in the $I\times I$ universality class, which we note is equivalent to the $B_2$ hypercubic theory with two scalars. One can solve \eqref{eq:betaone} explicitly to find the four non-trivial fixed points listed in Table \ref{tab:OB2}, where we note the remnant of the global symmetry group that is unbroken by the defect. Also included in the table are the number of stability matrix eigenvalues, $\kappa$, which are negative, corresponding to relevant $\phi^3$-type operators in the dCFT, or zero, which correspond to broken bulk symmetry currents. Interestingly, there are more fixed points here than when either bulk occupies the entirety of space. To see how these fixed points relate to the fixed points in either bulk, let us choose the following curve in the space of bulk couplings
\begin{equation}\label{eq:lambdachoiceOB2int}
    \lambda^1_{ijkl}=\lambda^2_{ijkl}=\alpha \lambda^{O(2)}_{ijkl}+(1-\alpha)\lambda^{B_2}_{ijkl}\,,
\end{equation}
which trivially interpolates between the $O(2)$ bulk for $\alpha=1$ and $I\times I$ for $\alpha=0$, passing through the $O(2)$--$I\times I$ interface for $\alpha=1/2$. For the bulk tensors we use (\ref{eq:lambdaON}) and (\ref{eq:lambdaBN}).
One can solve the beta function numerically for arbitrary values of $\alpha$, to find the families of solutions plotted in Figure \ref{fig:OC2int}. As $\alpha$ increases from zero, the two non-trivial fixed points in the $I\times I$ bulk move towards one another, eventually merging for $\alpha=1$ at one of the $O(2)$ fixed points, before splitting off again and moving off towards infinity for larger values of $\alpha$. The other two families of solutions only exist perturbatively for large enough $\alpha$, move off to infinity in the $\alpha\rightarrow0$ limit.

\begin{center}
\begin{longtable}{|c c c c|}\caption{Fixed points found for an $O(2)$--$I\times I$ RG interface, in terms of the invariants \eqref{eq:hinvariants}. $\kappa$ denotes the stability matrix eigenvalues.} \\
\hline
Symmetry & $h^2$ & $|h_i|^2$ & $\#\,\kappa<0$, =0 \\ [0.5ex]
 \hline
\endfirsthead

\multicolumn{4}{c}%
{{\bfseries \tablename\ \thetable{} -- continued from previous page}} \\
\hline
Symmetry & $h^2$ & $|h_i|^2$ & $\#\,\kappa<0$, =0 \\ [0.5ex]
 \hline
\endhead

\hline \multicolumn{4}{|r|}{{Continued on next page}} \\ \hline
\endfoot

\hline
\endlastfoot \label{tab:OB2}
$\mathbb{Z}_2$ & $\tfrac{1}{90}(225-\sqrt{33})$ & $\tfrac{1}{30}(57+7\sqrt{33})$ & 3, 0 \\
$\mathbb{Z}_2$ & $\tfrac{1}{90}(225+\sqrt{33})$ & $\tfrac{1}{30}(57-7\sqrt{33})$ & 2, 0 \\
$\mathbb{Z}_2$ & $4-\tfrac{8}{15\sqrt{3}}$ & $\tfrac{4}{15}(9+4\sqrt{3})$ & 4, 0 \\
$\mathbb{Z}_2$ & $4+\tfrac{8}{15\sqrt{3}}$ & $\tfrac{4}{15}(9-4\sqrt{3})$ & 3, 0 \\
\end{longtable}
\end{center}

For $N=3$ there are in principle $\frac{1}{2}\binom{9}{2}=18$ different pairings of bulks. The most interesting choices, however, are those involving the $O(3)$ and Cubic\footnote{We use `Cubic' with a capital `C' to refer to the $N=3$ (hyper)cubic bulk theory, not to be confused with the cubic coupling \eqref{eq:dacti} on the interface.} bulk, as these are both fully interacting fixed points, and are bulks which have been extensively studied in their own right. Though the number of equations in \eqref{eq:betaone} becomes too large to solve analytically, we can exhaustively search through the space of fixed points numerically, using the method outlined in \cite{Harribey:2024gjn}. As there now exists a non-trivial cubic interface in the free theory, cubic fixed points begin to appear even when we take one side of the interface to be free. For an $O(3)$--Free interface, we find one non-trivial fixed point, given in Table \ref{tab:O3}, while for a Cubic--Free interface we find two non-trivial fixed points, given in Table \ref{tab:B3}.

\begin{center}
\begin{longtable}{|c c c c|}\caption{Fixed points found for an $O(3)$--Free RG interface. $\kappa$ denotes the stability matrix eigenvalues.} \\
\hline
Symmetry & $h^2$ & $|h_i|^2$ & $\#\,\kappa<0$, =0 \\ [0.5ex]
 \hline
\endfirsthead

\multicolumn{4}{c}%
{{\bfseries \tablename\ \thetable{} -- continued from previous page}} \\
\hline
Symmetry & $h^2$ & $|h_i|^2$ & $\#\,\kappa<0$, =0 \\ [0.5ex]
 \hline
\endhead

\hline \multicolumn{4}{|r|}{{Continued on next page}} \\ \hline
\endfoot

\hline
\endlastfoot \label{tab:O3}
$O(2)$ & 31.2891 & 2.8499 & 7, 2 \\
\end{longtable}
\end{center}


\begin{center}
\begin{longtable}{|c c c c|}\caption{Fixed points found for a Cubic--Free RG interface. $\kappa$ denotes the stability matrix eigenvalues.} \\
\hline
Symmetry & $h^2$ & $|h_i|^2$ & $\#\,\kappa<0$, =0 \\ [0.5ex]
 \hline
\endfirsthead

\multicolumn{4}{c}%
{{\bfseries \tablename\ \thetable{} -- continued from previous page}} \\
\hline
Symmetry & $h^2$ & $|h_i|^2$ & $\#\,\kappa<0$, =0 \\ [0.5ex]
 \hline
\endhead

\hline \multicolumn{4}{|r|}{{Continued on next page}} \\ \hline
\endfoot

\hline
\endlastfoot \label{tab:B3}
$S_3$ & 31.455 & 2.7074 & 9, 0 \\
$\mathbb{Z}_2^2$ & 31.9182 & 2.8357 & 8, 0 \\
$D_4$ & 33.3348 & 3.2526 & 7, 0 \\
\end{longtable}
\end{center}

\begin{figure}[h]
\centering
\begin{tikzpicture}
\begin{axis}[
    xmin=-0, xmax=35,
    ymin=0, ymax=10,
    xlabel= $h^2$,
    ylabel= $|h_i|^2$,
    ylabel style={rotate=-90},
    title={Some RG Interfaces},
    legend pos = outer north east,
    reverse legend,
]
\addplot+[
    only marks,
    mark=*,
    mark options={color=black},
    mark size=2pt]
table{Data/O3Free.dat};
\addlegendentry{$O(3)$--Free interface}
\addplot+[
    only marks,
    mark=square*,
    mark options={color=black},
    mark size=2pt]
table{Data/B3Free.dat};
\addlegendentry{Cubic--Free interface}
\addplot+[
    only marks,
    mark=triangle*,
    mark options={color=black},
    mark size=2pt]
table{Data/O3B3.dat};
\addlegendentry{$O(3)$--Cubic interface}
\addplot+[
    only marks,
    mark=square*,
    mark options={color=Dark2-B,fill=Dark2-B},
    mark size=2pt]
table{Data/B3.dat};
\addlegendentry{Cubic bulk \cite{Harribey:2024gjn}}
\addplot+[
    only marks,
    mark=*,
    mark options={color=Dark2-A,fill=Dark2-A},
    mark size=2pt]
table{Data/O3.dat};
\addlegendentry{$O(3)$ bulk \cite{Harribey:2024gjn}}
\end{axis}
\end{tikzpicture}
    \caption{Fixed points found for an interface with cubic couplings. The values for the $O(3)$ and Cubic bulks were taken from \cite{Harribey:2024gjn}, and can equivalently be computed using \eqref{eq:betaone} by setting $\lambda^1_{ijkl}=\lambda^2_{ijkl}$. Note that not shown is the single fixed point for a free bulk, which is located at (72.5147, 6.1402).}
\label{fig:O3B3}
\end{figure}
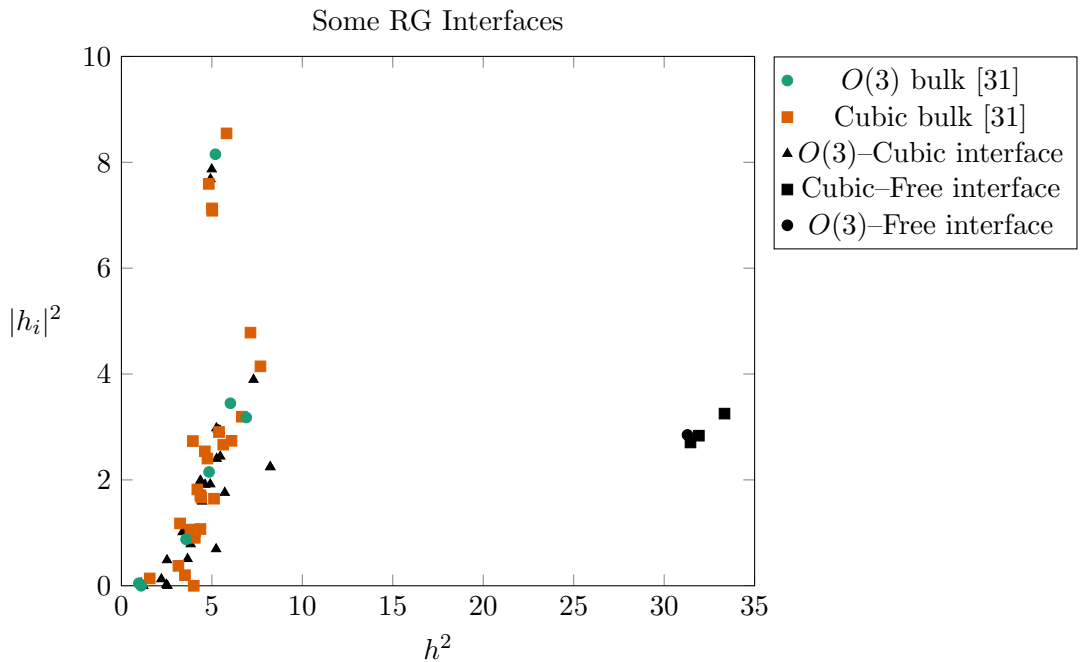

One can then choose $\lambda^1_{ijkl}=\lambda^{O(3)}_{ijkl}$ and $\lambda^2_{ijkl}=\lambda^{\text{Cubic}}_{ijkl}$ in \eqref{eq:betaone} to study cubic-deformed $O(3)$--Cubic RG interfaces\footnote{See \cite{Osborn:2017ucf} for a discussion of these bulk fixed points and their properties.}. We find a total of 29 fixed points in this system, listed in Table \ref{tab:OB3}, greater in number than the 26 fixed points found in the purely Cubic case. All of these fixed points, along with the $O(3)$ bulk and Cubic bulk fixed points, are displayed in Figure \ref{fig:O3B3}. There are too many fixed points to plot the families of solutions these points lie on, but were one to do so, the situation should be similar to the cases shown in Figures \ref{fig:OF2int}-\ref{fig:OC2int}.

\begin{center}
\begin{longtable}{|c c c c|}\caption{Fixed points found for an $O(3)$--Cubic RG interface. $\kappa$ denotes the stability matrix eigenvalues.} \\
\hline
Symmetry & $h^2$ & $|h_i|^2$ & $\#\,\kappa<0$, =0 \\ [0.5ex]
 \hline
\endfirsthead

\multicolumn{4}{c}%
{{\bfseries \tablename\ \thetable{} -- continued from previous page}} \\
\hline
Symmetry & $h^2$ & $|h_i|^2$ & $\#\,\kappa<0$, =0 \\ [0.5ex]
 \hline
\endhead

\hline \multicolumn{4}{|r|}{{Continued on next page}} \\ \hline
\endfoot

\hline
\endlastfoot \label{tab:OB3}
$\mathbb{Z}_2^2$ & 1.0927& 0.0512 & 4, 0 \\
$D_4$ & 1.2121 & 0 & 3, 0 \\
$\mathbb{Z}_2^2$ & 2.2099 & 0.1267 & 5, 0 \\
$S_3$ & 2.4861 & 0.0223 & 6, 0 \\
$\mathbb{Z}_2^2$ & 2.5212 & 0.4864 & 4, 0 \\
$S_4$ & 2.5455 & 0 & 7, 0 \\
$\mathbb{Z}_2^2$ & 3.3807 & 1.0147 & 6, 0 \\
$\mathbb{Z}_2^2$ & 3.6577 & 0.5082 & 5, 0 \\
$\mathbb{Z}_2$ & 3.828 & 0.7871 & 5, 0 \\
$\mathbb{Z}_2$ & 3.8324 & 0.9734 & 7, 0 \\
$\mathbb{Z}_2$ & 3.9711 & 0.9754 & 6, 0 \\
None & 4.3635 & 1.9879 & 6, 0 \\
$\mathbb{Z}_2$ & 4.3904 & 1.971 & 7, 0 \\
$\mathbb{Z}_2$ & 4.4631 & 1.5937 & 5, 0 \\
$\mathbb{Z}_2$ & 4.6267 & 1.9113 & 6, 0 \\
None & 4.9116 & 1.9195 & 7, 0 \\
$S_3$ & 4.925 & 7.684 & 8, 0 \\
$\mathbb{Z}_2^2$ & 4.9932 & 7.8686 & 9, 0 \\
$\mathbb{Z}_2^2$ & 5.2364 & 0.6944 & 6, 0 \\
$S_3$ & 5.2506 & 2.9774 & 6, 0 \\
$\mathbb{Z}_2$ & 5.265& 2.401 & 7, 0 \\
$S_3$ & 5.3352 & 2.9489 & 7, 0 \\
$\mathbb{Z}_2$ & 5.447 & 2.9328 & 7, 0 \\
$\mathbb{Z}_2$ & 5.4639 & 2.4447 & 8, 0 \\
$\mathbb{Z}_2$ & 5.7138 & 1.7591 & 6, 0 \\
$\mathbb{Z}_2$ & 6.8479 & 3.1688 & 8, 0 \\
$S_3$ & 7.3039 & 3.894 & 9, 0 \\
$\mathbb{Z}_2$ & 8.2258 & 2.2447 & 6, 0 \\
$\mathbb{Z}_2$ & 8.2317 & 2.2437 & 7, 0 \\
\end{longtable}
\end{center}

\subsection{Line Sub-Defect}
One can also consider integrating a relevant operator not over the entire interface, but instead along some submanifold embedded inside the interface. The simplest such submanifold will be a straight line, for which the defect action will be given by \eqref{eq:dactl}. For a single bulk, $\lambda^1_{ijkl}=\lambda^2_{ijkl}$, this line defect, also called the pinning field defect, has been extensively studied for a variety of bulk critical models\cite{Pannell:2023pwz}. The introduction of a non-zero coupling $h_i$ will induce a defect RG flow, governed by a beta function which can be determined by renormalizing the bulk one-point function $\langle\phi_i(x)\rangle$. As the counterterms are added directly to the action, it should not matter which side of the interface we choose to place this insertion. 
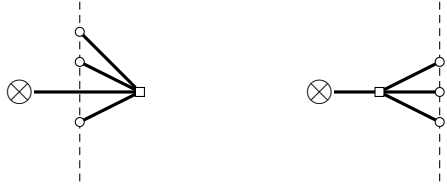
\begin{figure}[htbp]
    \centering
    \captionsetup[subfigure]{labelformat=empty}
    \subfloat[]{\begin{tikzpicture}[baseline=(vert_cent.base), square/.style={regular polygon,regular polygon sides=4},scale=0.4]
        \draw[densely dashed] (0,3)--(0,-3);
        \node[inner sep=0pt] at (-2,0) (l) {};
        \node at (2,0) [square, draw,fill=white,inner sep=1.2pt,outer sep=0pt]  (r) {};
        \draw[very thick] (l) to (r);
        \draw[very thick] (r) to (0,1);
        \draw[very thick] (r) to (0,-1);
        \draw[very thick] (r) to (0,2);
        \node at (0,1) [circle, draw,fill=white,inner sep=1.2pt,outer sep=0pt] {};
        \node at (0,2) [circle, draw,fill=white,inner sep=1.2pt,outer sep=0pt] {};
        \node at (0,-1) [circle, draw,fill=white,inner sep=1.2pt,outer sep=0pt] {};
        \node[fill=white,inner sep=0pt,outer sep=0pt, scale=1.27] at (-2,0) {$\otimes$};
        \node (vert_cent) at (current bounding box.center) {};
    \end{tikzpicture}}
    \hspace{2cm}
    \subfloat[]{\begin{tikzpicture}[baseline=(vert_cent.base), square/.style={regular polygon,regular polygon sides=4},scale=0.4]
        \draw[densely dashed] (0,3)--(0,-3);
        \node[inner sep=0pt] at (-4,0) (l) {};
        \node at (-2,0) [square, draw,fill=white,inner sep=1.2pt,outer sep=0pt]  (r) {};
        \draw[very thick] (l) to (r);
        \draw[very thick] (r) to (0,1);
        \draw[very thick] (r) to (0,0);
        \draw[very thick] (r) to (0,-1);
        \node at (0,1) [circle, draw,fill=white,inner sep=1.2pt,outer sep=0pt] {};
        \node at (0,0) [circle, draw,fill=white,inner sep=1.2pt,outer sep=0pt] {};
        \node at (0,-1) [circle, draw,fill=white,inner sep=1.2pt,outer sep=0pt] {};
        \node[fill=white,inner sep=0pt,outer sep=0pt, scale=1.27] at (-4,0) {$\otimes$};
        \node (vert_cent) at (current bounding box.center) {};
    \end{tikzpicture}}
    \caption{The two diagrams contributing to $\langle\phi_i(x)\rangle$ at $O(\varepsilon)$. The vertical dashed line represent the interface localised at $(\mathbf{x},0)$. We will make the choice that the insertion is on the left side of the interface, but this will not change the counterterms necessary for renormalisation.}
    \label{fig:linerenorm}
\end{figure}

At one loop the one-point function receives contributions from the two diagrams shown in Figure \ref{fig:linerenorm}, which differ only by which $\lambda^i$ they are proportional to. Let us consider first the diagram on the left, which is given by
\begin{equation}
    \frac{\lambda^1_{ijkl}h_jh_kh_l}{3!}C_\phi^4\int_0^\infty dz_\perp \int d\tau_z d^{d-2}\vec{z} d \tau_1d\tau_2d\tau_3\frac{1}{((x-z)^2)^{\frac{d-2}{2}}((\tau_1-z)^2)^{\frac{d-2}{2}}((\tau_2-z)^2)^{\frac{d-2}{2}}((\tau_3-z)^2)^{\frac{d-2}{2}}}
\end{equation}
As long as the bulk point isn't placed on the defect, its precise location will not matter from the point of view of the $1/\varepsilon$ pole, as translating it within the bulk will not add or subtract divergences. To simplify the computation, we then make the choice $x=(-|x_\perp|,\tau_x,\vec{0})$. It is then straightforward to perform all the integration save that over $z_\perp$ to find
\begin{equation}
     \frac{\lambda^1_{ijkl}h_jh_kh_l}{3!}\left(\frac{\Gamma\left(\frac{d-3}{2}\right)}{4\pi^{\frac{d-1}{2}}}\right)^4\frac{\pi^{\frac{d-2}{2}}\Gamma\left(\frac{3d-10}{2}\right)}{\Gamma(\frac{3d-9}{2})\Gamma\left(\frac{d-3}{2}\right)}\int_0^1d\alpha\int_{(1-\alpha)|x_\perp|}^\infty dz_\perp \frac{\alpha^{\frac{3d-9}{2}-1}(1-\alpha)^{\frac{d-3}{2}-1}}{(z_\perp^2+\alpha(1-\alpha)x_\perp^2)^{\frac{3d-10}{2}}}\,,
\end{equation}
where $\alpha$ is a Feynman parameter. To perform the remaining integral over $z_\perp$, we note that the integral can be split like 
\begin{equation}
    \int_{(1-\alpha)|x_\perp|}^\infty dz_\perp=\int_0^\infty dz_\perp-\int_0^{(1-\alpha)|x_\perp|}dz_\perp\,.
\end{equation}
The first integral is then manifestly exactly half of the integral one gets in the case of a single bulk, while the second integral is, in fact, finite in $d=4$. To see that this is the case, we note the following master integral
\begin{equation}
    \int_0^x dz \frac{(z^2)^a}{(z^2+\Delta)^b}=\frac{1}{2}\Delta^{-(b-a-\tfrac{1}{2})}B\left(\frac{x^2}{x^2+\Delta},a+\frac{1}{2},b-a-\frac{1}{2}\right)\,,
\end{equation}
where $B(z,a,b)$ is the incomplete beta function, which can be written in terms of hypergeometric functions by
\begin{equation}
    B(z,a,b)=\frac{z^a}{a}{_2}F_1(a,1-b\,;a+1\,;z)\,.
\end{equation}
Setting $d=4$ in the second integral, one finds that the hypergeometric simplifies and the integral becomes proportional to
\begin{equation}
    \int_0^1\sqrt{\frac{\alpha}{1-\alpha}}\text{arctanh}\,(1-\alpha)\approx0.832\,.
\end{equation}
As this is finite, and there are no divergences when $d$ is set to four in the gamma functions in the prefactor, the divergence is entirely contained in the first integral, which is half that for a complete bulk. That is,
\begin{equation}
    \begin{tikzpicture}[baseline=(vert_cent.base), square/.style={regular polygon,regular polygon sides=4},scale=0.4]
        \draw[densely dashed] (0,3)--(0,-3);
        \node[inner sep=0pt] at (-2,0) (l) {};
        \node at (2,0) [square, draw,fill=white,inner sep=1.2pt,outer sep=0pt]  (r) {};
        \draw[very thick] (l) to (r);
        \draw[very thick] (r) to (0,1);
        \draw[very thick] (r) to (0,-1);
        \draw[very thick] (r) to (0,2);
        \node at (0,1) [circle, draw,fill=white,inner sep=1.2pt,outer sep=0pt] {};
        \node at (0,2) [circle, draw,fill=white,inner sep=1.2pt,outer sep=0pt] {};
        \node at (0,-1) [circle, draw,fill=white,inner sep=1.2pt,outer sep=0pt] {};
        \node[fill=white,inner sep=0pt,outer sep=0pt, scale=1.27] at (-2,0) {$\otimes$};
        \node (vert_cent) at (current bounding box.center) {};
    \end{tikzpicture}=\frac{1}{1536\pi^3|x_\perp|\varepsilon}\lambda_{ijkl}h_j h_k h_l+\text{O}(\varepsilon^0)\,.
\end{equation}
Examining the second diagram, one sees that exactly the same thing will occur, so that we also have
\begin{equation}
    \begin{tikzpicture}[baseline=(vert_cent.base), square/.style={regular polygon,regular polygon sides=4},scale=0.4]
        \draw[densely dashed] (0,3)--(0,-3);
        \node[inner sep=0pt] at (-4,0) (l) {};
        \node at (-2,0) [square, draw,fill=white,inner sep=1.2pt,outer sep=0pt]  (r) {};
        \draw[very thick] (l) to (r);
        \draw[very thick] (r) to (0,1);
        \draw[very thick] (r) to (0,0);
        \draw[very thick] (r) to (0,-1);
        \node at (0,1) [circle, draw,fill=white,inner sep=1.2pt,outer sep=0pt] {};
        \node at (0,0) [circle, draw,fill=white,inner sep=1.2pt,outer sep=0pt] {};
        \node at (0,-1) [circle, draw,fill=white,inner sep=1.2pt,outer sep=0pt] {};
        \node[fill=white,inner sep=0pt,outer sep=0pt, scale=1.27] at (-4,0) {$\otimes$};
        \node (vert_cent) at (current bounding box.center) {};
    \end{tikzpicture}=\frac{1}{1536\pi^3|x_\perp|\varepsilon}\lambda_{ijkl}h_j h_k h_l+\text{O}(\varepsilon^0)\,.
\end{equation}
Absorbing these divergences into an appropriate counterterm for the coupling $h_i$, one then finds, after rescaling the couplings, the one-loop beta function
\begin{equation}\label{eq:betaline}
    \beta_i=-\frac{\varepsilon}{2}h_i+\frac{1}{6}\left(\frac{\lambda^1_{ijkl}+\lambda^2_{ijkl}}{2}\right)h_jh_kh_l\,.
\end{equation}
One sees that, as was the case of the cubic interaction placed on the entirety of the interface, at one loop the effect of the presence of two bulks on the line beta function is to simply replace $\lambda_{ijkl}\rightarrow\tfrac{\lambda^1_{ijkl}+\lambda^2_{ijkl}}{2}$.

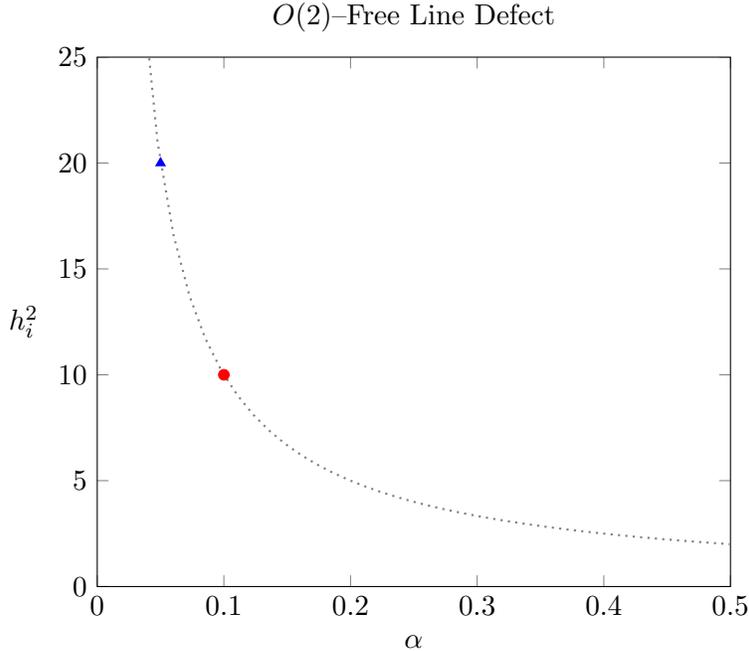
\begin{figure}[ht]
\centering
\begin{tikzpicture}
\begin{axis}[
    xmin=-0, xmax=0.5,
    ymin=0, ymax=25,
    xlabel= $\alpha$,
    ylabel= $h_i^2$,
    ylabel style={rotate=-90},
    title={$O(2)$--Free Line Defect}
]
\addplot[gray,dotted,thick,
domain=0.01:0.5,samples=40]
{1/x};
\addplot[only marks,mark=*,mark options={color=red},mark size=2pt] coordinates {(0.1, 10)};
\addplot[only marks,mark=triangle*,mark options={color=blue},mark size=2pt] coordinates {(0.05,20) };
\end{axis}
\end{tikzpicture}
    \caption{Family of solutions to the line defect beta function \eqref{eq:linebetanormal} for $O(2)$ symmetric values of $\lambda_{ijkl}$. The values of the coupling for the $O(2)$ bulk ($\alpha=\tfrac{1}{10}$), and the $O(2)$--Free interface ($\alpha=\tfrac{1}{20}$) are shown as the red circle and blue triangle respectively. One sees that as $\alpha$ decreases the fixed point moves towards strong coupling, explaining its non-existence in the free theory.}
\label{fig:OF2}
\end{figure}

\paragraph{Half-free bulk:}
Before continuing to the study of the fixed points of this beta function, let us first comment on the special case where one of the CFTs is taken to be free, $\lambda^2_{ijkl}=0$. In that case, the conformal lines will satisfy the equation
\begin{equation}
    -\frac{\varepsilon}{2}h^*_i+\frac{1}{12}\lambda^1_{ijkl}h^*_jh^*_kh^*_l=0\,,
\end{equation}
which is nearly the usual beta function for line defects in the bulk $\lambda^1$, except for an overall factor of 2 in the second term. Importantly, as there are no purely defect terms in the beta function, this factor of 2 can be completely absorbed in a rescaling of the defect coupling, so that conformal lines in a $\text{CFT}_1$--Free RG interface are completely classified by conformal lines in $\text{CFT}_1$. That is to say, suppose that $g^*_i$ is a conformal line in $\text{CFT}_1$. Then,
\begin{equation}
    -\frac{\varepsilon}{2}g^*_i+\frac{1}{6}\lambda^1_{ijkl}g^*_jg^*_kg^*_l=0\,.
    \label{eq:linebetanormal}
\end{equation}
Defining $h^*_i=\tfrac{1}{\sqrt{2}}g^*_i$, one sees that
\begin{equation}
    0=\frac{1}{\sqrt{2}}\beta^g_i(h^*)=-\frac{\varepsilon}{2}h^*_i+\frac{1}{12}\lambda^1_{ijkl}h^*_jh^*_kh^*_l=\beta^h_i(h^*)\,.
\end{equation}
Thus, there is a simple one-to-one mapping between conformal lines in a complete bulk, and conformal lines where half of the space is free.
One sees that this argument can be easily generalised to rescaling $\lambda^1_{ijkl}$ by any number, so that $h_i^*$ and $g_i^*$ should be connected by a continuous family of solutions satisfying \eqref{eq:linebetanormal} with $\lambda_{ijkl}(\alpha)=\alpha\lambda^1_{ijkl}$ for all values of $\alpha$. For instance, consider the following curve in the space of $\lambda_{ijkl}$:
\begin{equation}
    \lambda_{ijkl}(\alpha)=\alpha(\delta_{ij}\delta_{kl}+\text{Perms.})\,,
\end{equation}
which for $\alpha=\tfrac{\varepsilon}{N+8}$ gives the $O(N)$ fixed point, and for $\alpha=\tfrac{\varepsilon}{2N+16}$ will reproduce the $O(N)$--Free interface. One can straightforwardly solve \eqref{eq:linebetanormal} for these values of $\lambda_{ijkl}$ to obtain the family of solutions
\begin{equation}
    (h^*)^2=\frac{\varepsilon}{\alpha}\,.
\end{equation}
For the $O(N)$--Free interface, this becomes
\begin{equation}\label{eq:IsingFreepinningdefect}
    (h^*)^2=2(N+8)\,.
\end{equation}
For $N=2$ this family of solutions, with the $O(2)$ bulk and $O(2)$--Free interface highlighted, is shown in Figure \ref{fig:OF2}. One can also immediately see that these fixed points will retain the stability properties of their parent conformal lines in the single bulk. The stability matrix of a fixed point $h^*_i$ in a $\text{CFT}_1$--Free RG interface is given by
\begin{equation}
    S_{ij}=-\frac{\varepsilon}{2}\delta_{ij}+\frac{1}{4}\lambda^1_{ijkl}h^*_kh^*_l=-\frac{\varepsilon}{2}\delta_{ij}+\frac{1}{2}\lambda^1_{ijkl}g^*_kg^*_l\,,
\end{equation}
which we recognise as the stability matrix of the corresponding fixed point for the purely $\lambda^1$ bulk. It is important to note that the simplicity of this map, just replacing $g^*_i$ by $\tfrac{1}{\sqrt{2}}g^*_i$, does not have to hold at higher orders, as the two-loop terms in the beta function do not scale uniformly with $h_i$. Nevertheless, as higher-order corrections to the fixed point are completely determined by the one-loop value, the existence of a one-to-one map, and the invariance of stability properties under this map, will persist to higher-loop orders.

Recently, the authors of \cite{Lanzetta:2025xfw} were able to locate the pinning field defect for an Ising bulk using the numerical conformal bootstrap. As the Ising--Free RG interface is conformal, and has the same $\mathbb{Z}_2$ symmetry as the Ising model, it is worthwhile to comment on why \eqref{eq:IsingFreepinningdefect} was not identified in \cite{Anataichuk:2025zoq}. In their bootstrap setup, they include four-point functions containing the two lowest scalar operators in the Ising model, $\sigma$ and $\varepsilon$, in order to tell the bootstrap that the one-dimensional CFT is embedded within an Ising bulk. While $\sigma$ and $\varepsilon$ will exist as operators on one side of the interface, as we have placed the line within the interface, one would instead have to use $\hat{\sigma}$ and $\hat{\varepsilon}$, the two lowest lying scalars in the interface CFT. Following the techniques outlined in \cite{Gliozzi:2015qsa}, it should be possible to then use the techniques of \cite{Lanzetta:2025xfw} to identify the conformal lines found in this section.

\paragraph{Non-free bulk:}
The simplicity of solutions to \eqref{eq:betaline} breaks down when both bulks are taken to be interacting, as it is no longer possible to absorb the averaging of $\lambda^1$ and $\lambda^2$ by an overall rescaling of $h_i$. Instead, we find a considerably richer space of fixed points, uncovering a vast network of solution families. As illustrative examples of the types of new fixed points which can arise, let us consider two different RG interfaces: an interface between the $O(N)$ CFT and the hypercubic CFT, and an interface between a hypercubic CFT and a hypertetrahedral CFT. Fixed points in a hypercubic bulk were worked out in \cite{Pannell:2023pwz}, while a partial classification of those in a hypertetrahedral bulk can be found in \cite{Pannell:2024hbu}. 

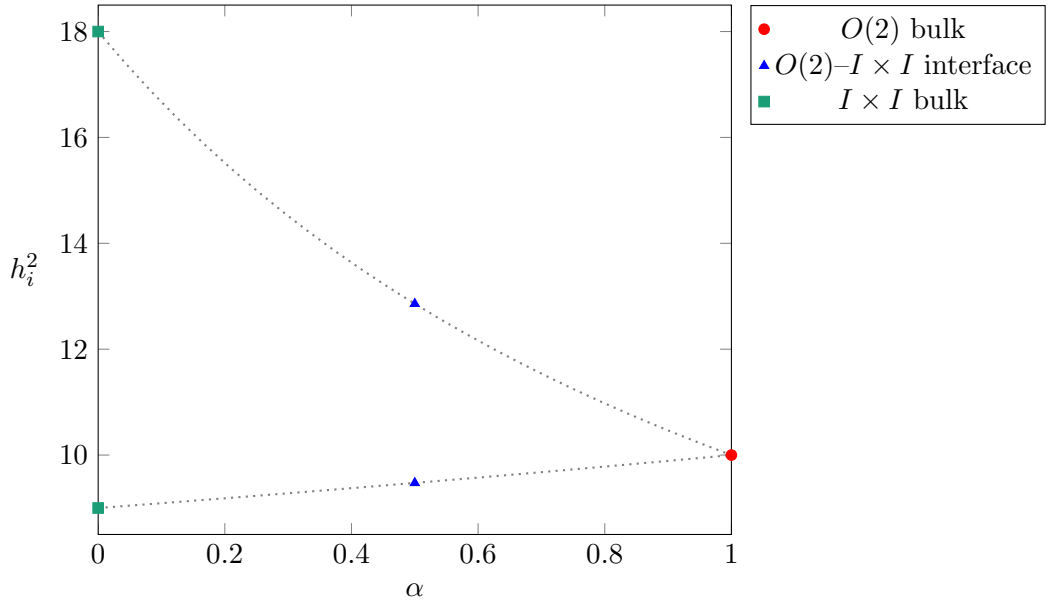
\begin{figure}[h]
\centering
\begin{tikzpicture}
\begin{axis}[
    xmin=-0, xmax=1,
    ymin=8.5, ymax=18.5,
    xlabel= $\alpha$,
    ylabel= $h_i^2$,
    ylabel style={rotate=-90},
    title={$O(2)$--$I\times I$ Line Defect},
    legend pos=outer north east
]
\addplot[only marks,mark=*,mark options={color=red},mark size=2pt] coordinates {(1, 10)};
\addlegendentry{$O(2)$ bulk};
\addplot[only marks,mark=triangle*,mark options={color=blue},mark size=2pt] coordinates {(0.5, 12.8571) (0.5, 9.47368)};
\addlegendentry{$O(2)$--$I\times I$ interface};
\addplot[only marks,mark=square*,mark options={color=Dark2-A},mark size=2pt] coordinates {(0,9) (0,18)};
\addlegendentry{$I\times I$ bulk};
\addplot[gray,dotted,thick,
domain=0:1,samples=25]
{90/(5+4*x)};
\addplot[gray,dotted,thick,
domain=0:1,samples=25]
{-90/(-10+x)};
\end{axis}
\end{tikzpicture}
    \caption{Families of solutions to the line defect beta function \eqref{eq:linebetanormal}, where $\lambda_{ijkl}$ takes values along the curve \eqref{eq:ONCubiccurve} for $N=2$. The families are shown as dotted lines, while the physical conformal defects are highlighted. It is important to note that the stability properties of the defect are preserved for all $0\leq \alpha<1$, so that the upper family of solutions is stable, while the lower family is unstable.}
\label{fig:OC2}
\end{figure}

To study lines in an $O(N)$ bulk, hypercubic bulk, and $O(N)$--hypercubic RG interface simultaneously, let us choose the following parametrisation of the bulk tensor
\begin{equation}\label{eq:ONCubiccurve}
    \lambda^1_{ijkl}=\lambda^2_{ijkl}=\alpha\lambda^{O(N)}_{ijkl}+(1-\alpha)\lambda^{B_N}_{ijkl}=\frac{(2\alpha(N-4)+N+8)\varepsilon}{3N(N+8)}(\delta_{ij}\delta_{kl}+\text{Perms.})+\frac{(1-\alpha)(N-4)\varepsilon}{3N}\delta_{ijkl}\,.
\end{equation}
The analysis of fixed points follows along very similar lines to that found in \cite{Pannell:2023pwz} for the purely hypercubic bulk, though now with added $\alpha$ dependence. Substituting this into \eqref{eq:betaline}, one finds that fixed points satisfy
\begin{equation}
    \left(-\frac{1}{2}+\frac{2\alpha(N-4)+N+8}{6N(N+8)}h^2+\frac{(1-\alpha)(N-4)}{18N}h_i^2\right)\varepsilon h_i=0\,.
\end{equation}
Clearly, for a given $i$, either $h_i=0$, or
\begin{equation}
    h_i^2=\frac{18N}{(1-\alpha)(N-4)}\left(\frac{1}{2}-\frac{2\alpha(N-4)+N+8}{6N(N+8)}h^2\right)\,.
\end{equation}
As the right hand side is the same regardless of which value of $i$ we choose, one sees that all solutions will have some number, $n$, of non-zero, equal magnitude couplings. As solutions related by elements of $B_N$ are equivalent, we can consistently choose the first $n$ couplings to be non-zero. Then, one finds the $n$ non-trivial families of solutions
\begin{equation}
    h_i^2=\begin{cases}
    \frac{9N(N+8)}{(N-4)(N+8)(1-\alpha)+3n(2\alpha(N-4)+N+8)}\qquad&i\leq n\\
    0&i>n
    \end{cases}
\end{equation}
which one can check agrees with the usual $O(N)$ and hypercubic fixed points when $\alpha$ is set to 1 and 0 respectively. For $\alpha=1/2$, we find the $n$ conformal line defects inside the $O(N)$--Hypercubic RG interface,
\begin{equation}
    h_i^2=\begin{cases}
    \frac{18N(N+8)}{(N-4)(N+8)+12n(N+2)}\qquad&i\leq n\\
    0&i>n
    \end{cases}\,.
\end{equation}
For $0\leq \alpha<1$, the stability properties of the above families is constant. To see this, note that the stability matrix at a given fixed point is
\begin{equation}
    \partial_i\beta_j|_{h=h^*}=\varepsilon\delta_{ij}\left(-\frac{1}{2}+\frac{2\alpha(N-4)+N+8}{6N(N+8)}(h^*)^2+\frac{(1-\alpha)(N-4)}{6N}(h_i^*)^2\right)+\frac{2\alpha(N-4)+N+8}{3N(N+8)}h^*_i h^*_j\,.
\end{equation}
Following the logic in \cite{Pannell:2023pwz}, we divide this into block-diagonal form, as
\begin{equation}
    \partial_i\beta_j=\begin{pmatrix}
        \textbf{P} & \textbf{0} \\ \textbf{0} & \textbf{Q}
    \end{pmatrix}\,,
\end{equation}
where
\begin{equation}
\begin{split}
    \textbf{P}&=\frac{\varepsilon}{1+3n \frac{2\alpha(N-4)+N+8}{(N-4)(N+8)(1-\alpha)}}\textbf{Id}_{n}+\frac{3\varepsilon}{3n+\frac{(N-4)(N+8)(1-\alpha)}{2\alpha(N-4)+N+8}}\begin{pmatrix}
1 & 1 & \cdots & 1\\
1 & 1 & \cdots & 1\\
\vdots & \vdots  & \ddots & \vdots\\
1 & 1 & \cdots & 1\\
\end{pmatrix}\,,\\
    \textbf{Q}&=-\frac{\varepsilon}{2+6n\frac{2\alpha(N-4)+N+8}{(N-4)(N+8)(1-\alpha)}}\textbf{Id}_{N-n}\,,
\end{split}
\end{equation}
and $\textbf{Id}_m$ is the $m\times m$ identity matrix. Naturally, this reproduces the correct stability matrices for the $O(N)$ and hypercubic pinning field defects, for the appropriate values of $\alpha$. One immediately sees that for $N>4$ and $0\leq \alpha<1$, \textbf{Q} will be negative definite, so that all fixed points with $n<N$ will be unstable. For $n=N$, the stability matrix has one eigenvalue with eigenvalue 1, and then $N-1$ eigenvalues
\begin{equation}
    \kappa=\frac{\varepsilon}{1+3N\frac{2\alpha(N-4)+N+8}{(N-4)(N+8)(1-\alpha)}}\,,
\end{equation}
which is positive for $N>4$ and $0\leq \alpha<1$. $N=4$ is a special case, as the two bulks are the same, so the $\alpha$-dependence drops out, and the single remaining fixed point is simply the $O(4)$ pinning field defect. For $N<4$, one may check explicitly that it is instead the $n=1$ point which is stable, while all others are unstable, and that this holds for all solutions with $0\leq\alpha<1$. One sees that in all cases, the stability property is shared by all points in a given solution family, so that the conformal line in the $O(N)$--Hypercubic RG interface inherits the stability properties of its parent solution in the hypercubic bulk. The case for $N=2$ is shown in Figure \ref{fig:OC2}, where we note that the $B_2$ fixed point is isomorphic to two decoupled Ising models.

\begin{figure}[H]
\centering
\begin{tikzpicture}
\begin{axis}[
    xmin=-0, xmax=1,
    ymin=9.5, ymax=18.5,
    xlabel= $\alpha$,
    ylabel= $h_i^2$,
    ylabel style={rotate=-90},
    title={$B_5$--$T_5$ Line Defect},
    legend pos=outer north east,
    reverse legend
]
\addplot+[
    only marks,
    mark=square,
    mark options={color=gray},
    mark size=0.1pt]
table{Data/cubtetplot.dat};
\addplot+[
    only marks,
    mark=square*,
    mark options={color=Dark2-A},
    mark size=2pt]
table{Data/cubtetl.dat};
\addlegendentry{$T_5$ bulk};
\addplot+[
    only marks,
    mark=triangle*,
    mark options={color=blue},
    mark size=2pt]
table{Data/cubtetm.dat};
\addlegendentry{$B_5$--$T_5$ interface};
\addplot+[
    only marks,
    mark=*,
    mark options={color=red},
    mark size=2pt]
table{Data/cubtetr.dat};
\addlegendentry{$B_5$ bulk};
\end{axis}
\end{tikzpicture}
    \caption{Families of solutions to the line defect beta function \eqref{eq:linebetanormal}, where $\lambda_{ijkl}$ takes values along the curve \eqref{eq:CTcurve} for $N=5$. The families are shown as dotted lines, while the physical conformal defects are highlighted. Note the increased number of physical conformal defects present for the interface, compared to either of the bulks; due to the reduced $\mathbb{Z}^2_2$ symmetry of the interface theory, there are more distinct solutions.}
\label{fig:CT5}
\end{figure}
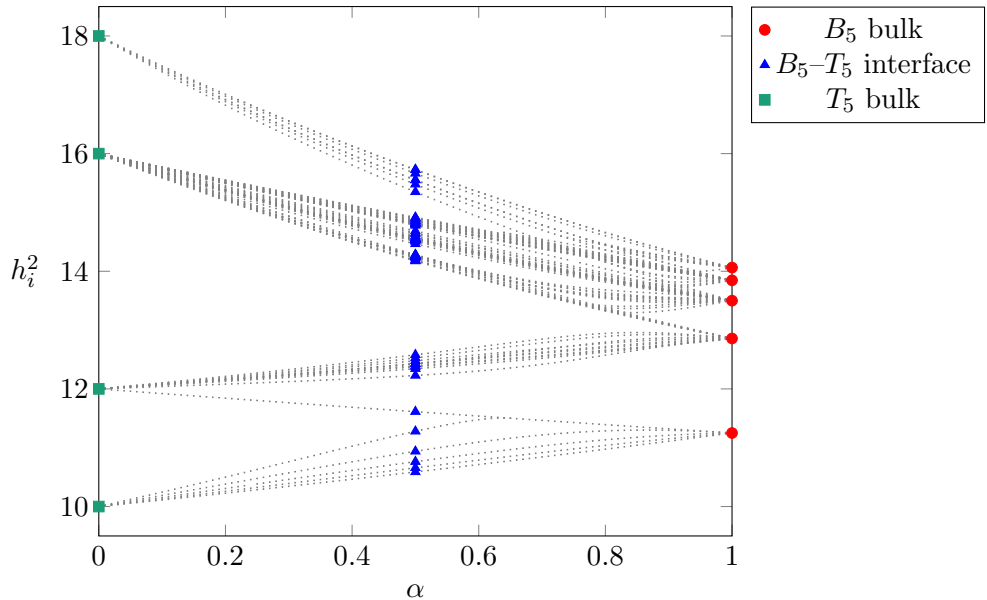

Let us now look at an interface between a hypercubic CFT and a hypertetrahedral CFT. To study the RG interface system, we will again be interested in identifying families of solutions which connect the conformal lines in the hypercubic model, with conformal lines in the hypertetrahedral model. To that end, we consider the curve
\begin{equation}\label{eq:CTcurve}
    \lambda^1_{ijkl}=\lambda^2_{ijkl}=\alpha \lambda^{B_N}_{ijkl}+(1-\alpha)\lambda^{T_N^-}_{ijkl}\,,
\end{equation}
where for the bulk tensors we use the values (\ref{eq:lambdaBN}) and (\ref{eq:lambdaTN}). The form of $\lambda^{T_N}_{ijkl}$ is rather complicated, which makes analysis of conformal lines in this system difficult, and we are forced to work numerically. The situation for $N=5$, where $T_N^+$ and $T_N^-$ are isomorphic fixed points, is shown in Figure \ref{fig:CT5}. One sees that unlike for the $O(N)$--Hypercubic interface, where the number of solution families was equal to the number of fixed points in the purely hypercubic case, a wealth of solutions appears as soon as one moves away from one of the critical bulk tensors with as many as 81 solutions appearing for generic $\alpha$. One sees that if one were to only consider conformal bulks from the beginning, the presence of a large number of solutions would be missed entirely. We find 68 fixed points at the RG interface, for $\alpha=1/2$, of which 7 are totally stable fixed points. It seems that once again, the stability properties are constant along the solution families, as all of these seven stable points lie on trajectories which begin and end at the unique stable fixed point on either end. 

The existence of an extremely fine structure in the solutions in this case, as opposed to the other cases considered in this section, comes about from a mismatch in the symmetries on either side of the interface. As $B_N\subset O(N)$, the symmetry of the $O(N)$--Hypercubic interface will be $B_N$, the same as a purely hypercubic bulk. Consequently, the symmetry breaking patterns induced by the addition of the line defect will be the same, and the number of solution families will equal the number of fixed points in the purely hypercubic bulk. As $B_N\not\subset T_N$ and $T_N\not\subset B_N$, placing hypercubic and a hypertetrahedral bulks opposite one another will induce significant explicit global symmetry breaking, in the case of $N=5$ down to the Klein four-group $\mathbb{Z}_2^2$. From the point of view of the line defect beta function, the line has been placed in a background with considerably less symmetry than is present for any conformal value of $\lambda_{ijkl}$, and there will thus be many more distinct solutions. One sees that by choosing different combinations of the CFTs on either side of the interface, one can gain access to conformal lines with considerably less symmetry than would typically be present for a single bulk.

Here, we find that of the 68 fixed points in the $B_5$--$T_5$ interface, 39 have a $\mathbb{Z}_2$ symmetry remaining, while 29 have no global symmetry remaining, of which three are totally RG stable. This is in stark contrast to the situation for line defects within a single bulk, where no conformal lines without any global symmetry have been found.

\subsection{Surface Sub-Defect}
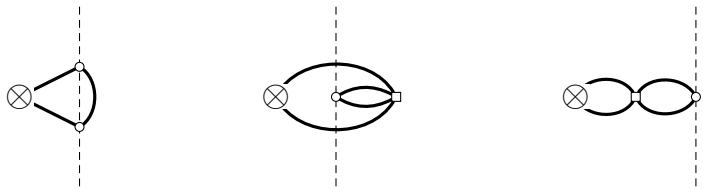
\begin{figure}[htbp]
    \centering
    \captionsetup[subfigure]{labelformat=empty}
    \subfloat[]{\begin{tikzpicture}[baseline=(vert_cent.base), square/.style={regular polygon,regular polygon sides=4},scale=0.4]
        \draw[densely dashed] (0,3)--(0,-3);
        \node[inner sep=0pt] at (-2,0) (l) {};
        \draw[very thick] (l) to (0,1);
        \draw[very thick] (l) to (0,-1);
        \draw[very thick] (0,1) to[bend left=60] (0,-1);
        \node at (0,1) [circle, draw,fill=white,inner sep=1.2pt,outer sep=0pt] {};
        \node at (0,-1) [circle, draw,fill=white,inner sep=1.2pt,outer sep=0pt] {};
        \node[fill=white,inner sep=0pt,outer sep=0pt, scale=1.27] at (-2,0) {$\otimes$};
        \node (vert_cent) at (current bounding box.center) {};
    \end{tikzpicture}}\hspace{2cm}
    \subfloat[]{\begin{tikzpicture}[baseline=(vert_cent.base), square/.style={regular polygon,regular polygon sides=4},scale=0.4]
        \draw[densely dashed] (0,3)--(0,-3);
        \node[inner sep=0pt] at (-2,0) (l) {};
        \node at (2,0) [square, draw,fill=white,inner sep=1.2pt,outer sep=0pt]  (r) {};
        \draw[very thick] (l) to[bend left=60] (r);
        \draw[very thick] (l) to[bend right=60] (r);
        \draw[very thick] (r) to[bend left=30] (0,0);
        \draw[very thick] (r) to[bend right=30] (0,0);
        \node at (0,0) [circle, draw,fill=white,inner sep=1.2pt,outer sep=0pt] {};
        \node[fill=white,inner sep=0pt,outer sep=0pt, scale=1.27] at (-2,0) {$\otimes$};
        \node (vert_cent) at (current bounding box.center) {};
    \end{tikzpicture}}
    \hspace{2cm}
    \subfloat[]{\begin{tikzpicture}[baseline=(vert_cent.base), square/.style={regular polygon,regular polygon sides=4},scale=0.4]
        \draw[densely dashed] (0,3)--(0,-3);
        \node[inner sep=0pt] at (-4,0) (l) {};
        \node at (-2,0) [square, draw,fill=white,inner sep=1.2pt,outer sep=0pt]  (r) {};
        \draw[very thick] (l) to[bend left=60] (r);
        \draw[very thick] (l) to[bend right=60] (r);
        \draw[very thick] (r) to[bend left=60] (0,0);
        \draw[very thick] (r) to[bend right=60] (0,0);
        \node at (0,0) [circle, draw,fill=white,inner sep=1.2pt,outer sep=0pt] {};
        \node[fill=white,inner sep=0pt,outer sep=0pt, scale=1.27] at (-4,0) {$\otimes$};
        \node (vert_cent) at (current bounding box.center) {};
    \end{tikzpicture}}
    \caption{The three diagrams contributing to $\langle\phi_i(x)^2\rangle$ at $O(\varepsilon)$. The vertical dashed line represent the interface localised at $(\mathbf{x},0)$. We will make the choice that the insertion is on the left side of the interface, but this will not change the counterterms necessary for renormalisation.}
    \label{fig:surfrenorm}
\end{figure}

Finally, we can consider constructing a defect on a two-dimensional submanifold of the RG interface, with the defect action given by \eqref{eq:dacts}. The beta function for $h_{ij}$ can be determined by renormalizing the bulk two-point function $\langle\phi^2_{ij}(x)\rangle$, where we can again choose to insert the operator on either side of the interface. For a single bulk, the fixed points of this beta function have been extensively studied for a variety of bulks in \cite{Anataichuk:2025zoq}. At one loop, there will be three diagrams contributing to the two-point function, shown in Figure \ref{fig:surfrenorm}.

The first of these diagrams is a purely defect contribution, and will not know about the presence of the RG interface, so that we can borrow the result from \cite{Trepanier:2023tvb,Anataichuk:2025zoq}:
\begin{equation}
    \begin{tikzpicture}[baseline=(vert_cent.base), square/.style={regular polygon,regular polygon sides=4},scale=0.4]
        \draw[densely dashed] (0,3)--(0,-3);
        \node[inner sep=0pt] at (-2,0) (l) {};
        \draw[very thick] (l) to (0,1);
        \draw[very thick] (l) to (0,-1);
        \draw[very thick] (0,1) to[bend left=60] (0,-1);
        \node at (0,1) [circle, draw,fill=white,inner sep=1.2pt,outer sep=0pt] {};
        \node at (0,-1) [circle, draw,fill=white,inner sep=1.2pt,outer sep=0pt] {};
        \node[fill=white,inner sep=0pt,outer sep=0pt, scale=1.27] at (-2,0) {$\otimes$};
        \node (vert_cent) at (current bounding box.center) {};
    \end{tikzpicture}=\frac{1}{32\pi^4x_\perp^2\varepsilon}h_{ik}h_{jk}+\text{O}(\varepsilon^0)\,.
\end{equation} 
For the other two diagrams, we note that because they are linear in the coupling, their combined contribution to the beta function will be via the anomalous dimension of the operator $\phi^2_{ij}$,
\begin{equation}
\beta_{ij}\supset \gamma^{\phi^2}_{ij,kl}h_{kl}\,.    
\end{equation}
As the surface is inserted within the RG interface, the $\phi^2$ operator in the integral is not an operator in either bulk CFT, but instead the $\phi^2$ operator in the interface CFT. It was shown in \cite{Gliozzi:2015qsa} that at one loop the anomalous dimension on the interface of scalar operators such as $\phi^2$ is simply the average of the anomalous dimensions of $\phi^2$ on either side of the interface. Thus, the linear term in the beta function will be
\begin{equation}
\beta_{ij}\supset \frac{\lambda^1_{ijkl}+\lambda^2_{ijkl}}{2}h_{kl}\,.\end{equation} 
This term can also be derived by noting that as the dependence on $\lambda_1$ and $\lambda_2$ must be linear and symmetric, and we must reproduce the usual surface defect beta function when $\lambda_1=\lambda_2$, one can simply replace $\lambda_{ijkl}\rightarrow\tfrac{\lambda^1_{ijkl}+\lambda^2_{ijkl}}{2}$ in the one-loop beta function. Putting this together, we find
\begin{equation}\label{eq:surfbeta}
    \beta_{ij}=-\varepsilon h_{ij}+\frac{\lambda^1_{ijkl}+\lambda^2_{ijkl}}{2}h_{kl}+h_{ik}h_{jk}\,,
\end{equation}
for a surface within an RG interface. Note that once again the couplings have been rescaled.

\paragraph{Half-free bulk:} One can again search for fixed points of this beta function. When $\lambda^2$ is free, analysis of the fixed points will be very similar to the case of a single bulk\cite{Anataichuk:2025zoq}, though the additional factor of $1/2$ in front of the bulk term will alter the precise numerical factors. The important point is that there will be a one-to-one correspondence between single $\text{CFT}_1$ bulk fixed points and $\text{CFT}_1$--Free RG interface fixed points, as there was for line defects, though it is no longer possible to construct them via a simple rescaling argument. 

When $\text{CFT}_1$ is the $O(N)$ model, we can diagonalise $h_{ij}$ to diagonalise the beta function
\begin{equation}
\beta_{ij}=\left(-\varepsilon h_{ii}+\frac{\varepsilon}{2(N+8)}(h_{kk}+2h_{ii})+h_{ii}^2\right)\delta_{ij}\qquad\qquad \text{No sum over }i\,.
\end{equation}
Fixed points are classified by splitting the couplings $h_{ii}$ into two groups, with
\begin{equation}\label{eq:surfsplit}
    h_{ii}=\begin{cases}
        \mathfrak{h}\,,\qquad &i\leq p\\
        \mathfrak{g}\,, &i>p
    \end{cases}\,,
\end{equation}
for $0\leq p\leq N$. One then finds the $N+1$ solutions
\begin{equation}
    \mathfrak{h}=\frac{3N-2p+14+\sqrt{4p^2-4pN+(N+14)^2}}{4(N+8)}\varepsilon\,,\qquad \mathfrak{g}=\frac{2p-N+14-\sqrt{4p^2-4pN+(N+14)^2}}{4(N+8)}\varepsilon\,.
\end{equation}

When $\text{CFT}_1$ is the hypercubic model, we can again simultaneously diagonalise $h_{ij}$ and $\beta_{ij}$ to obtain
\begin{equation}
\beta_{ij}=\left(-\varepsilon h_{ii}+\frac{\varepsilon}{6N}(h_{kk}+2h_{ii})+\frac{(N-4)\varepsilon}{6N}h_{ii}+h_{ii}^2\right)\delta_{ij}\qquad\qquad \text{No sum over }i\,.
\end{equation}
We can again classify the fixed points by splitting the couplings as in \eqref{eq:surfsplit}, finding the $N+1$ solutions
\begin{equation}
    \mathfrak{h}=\frac{3N-p+1+\sqrt{p^2-p N+(2N+1)^2}}{6N}\varepsilon\,,\qquad\mathfrak{g}=\frac{2N+p+1-\sqrt{p^2-p N+(2N+1)^2}}{6N}\varepsilon\,.
\end{equation}
It is important to comment, however, that this is not an exhaustive list of conformal surfaces in the hypercubic model. While one can always diagonalise $h_{ij}$ by an $O(N)$ rotation, it is not necessarily possible to do this with an element of $B_N$, and thus such a rotation would alter the form of $\lambda_{ijkl}^{B_N}$. For instance, by solving the beta function directly for $N=2$ we find one additional non-diagonal solution, and for $N=3$ we find five additional non-diagonal solutions.

Finally, we consider the case when $\text{CFT}_1$ is the hypertetrahedral model, focusing on the $T_N^-$ for simplicity. Here, diagonalizing $h_{ij}$ does not diagonalise $\beta_{ij}$, but the off-diagonal terms are remarkably simple\cite{Anataichuk:2025zoq}, and will simply set
\begin{equation}\label{eq:surftetconddiag}
    h_{ii}=\mathfrak{h}\qquad\forall i\leq N-1\,.
\end{equation}
One then finds the three fixed points
\begin{equation}
\begin{aligned}
    \mathfrak{h}&=h_{NN}=\frac{2(N+4)}{3(N+3)}\varepsilon\,, &&{}\\
    \mathfrak{h}&=\frac{N (2 N+9)+1-\sqrt{ P_N}}{6
   N (N+3)}\varepsilon\,, & \qquad  h_{NN} &=\frac{N (3 N+8)-1+\sqrt{P_N}}{6
   N (N+3)}\varepsilon\,, \\
    \mathfrak{h}&=\frac{N (2 N+9)+1+\sqrt{P_N}}{6
   N (N+3)}\varepsilon\,, &  h_{NN}&=\frac{N (3 N+8)-1-\sqrt{P_N}}{6
   N (N+3)}\varepsilon\,,
\end{aligned}
\end{equation}
where $P_N=N^2(4 N^2+31N+63)+N+1$. As with the hypercubic bulk, this will not exhaust the space of conformal surfaces, and there will exist non-diagonal fixed points which cannot be diagonalised without rotating $\lambda_{ijkl}^{T_N}$. 

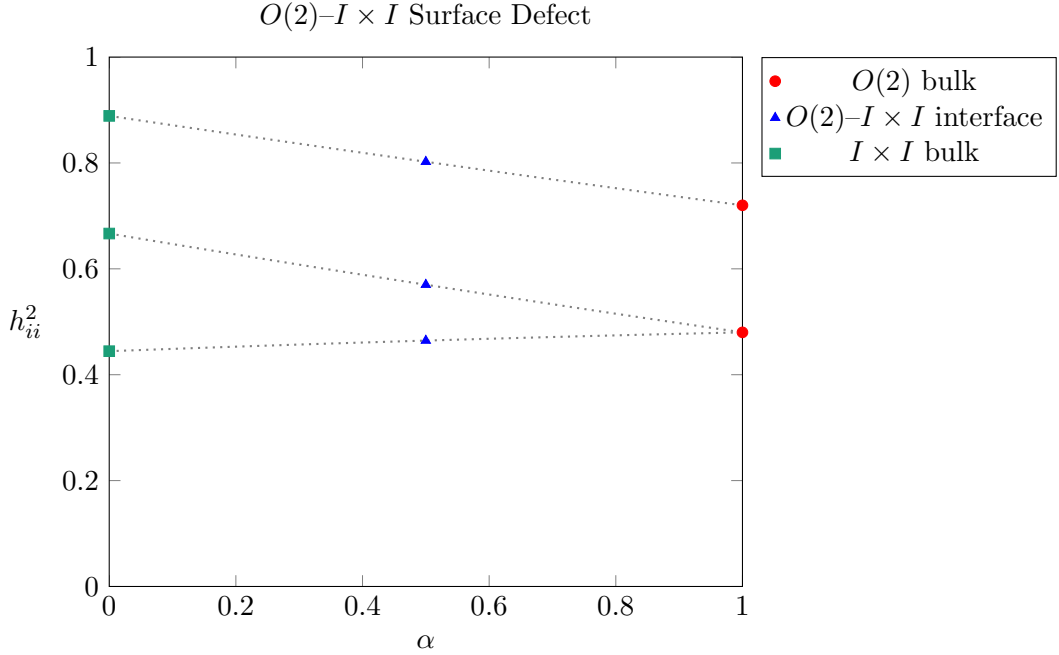
\begin{figure}[h]
\centering
\begin{tikzpicture}
\begin{axis}[
    xmin=-0, xmax=1,
    ymin=0, ymax=1,
    xlabel= $\alpha$,
    ylabel= $h_{ii}^2$,
    ylabel style={rotate=-90},
    title={$O(2)$--$I\times I$ Surface Defect},
    legend pos=outer north east,
    reverse legend
]
\addplot[gray,dotted,thick,
domain=0:1,samples=25]
{-2*(-50-5*x+x^2)/225};
\addplot[gray,dotted,thick,
domain=0:1,samples=25]
{(50-15*x+x^2)/75};
\addplot[gray,dotted,thick,
domain=0:1,samples=25]
{2*(-10+x)^2/225};
\addplot[only marks,mark=square*,mark options={color=Dark2-A},mark size=2pt] coordinates {(0,0.88889) (0,0.66667) (0,0.44444)};
\addlegendentry{$I\times I$ bulk};
\addplot[only marks,mark=triangle*,mark options={color=blue},mark size=2pt] coordinates {(0.5, 0.80222) (0.5, 0.57) (0.5, 0.46444)};
\addlegendentry{$O(2)$--$I\times I$ interface};
\addplot[only marks,mark=*,mark options={color=red},mark size=2pt] coordinates {(1., 0.72) (1.,0.48)};
\addlegendentry{$O(2)$ bulk};
\end{axis}
\end{tikzpicture}
    \caption{Families of solutions to the surface defect beta function \eqref{eq:surfbeta}, where $\lambda_{ijkl}$ takes values along the curve \eqref{eq:ONCubiccurve} with $N=2$. The families are shown as dotted lines, while the physical conformal defects are highlighted. Note the additional non-diagonal family of solutions, with $g^2=\tfrac{4}{9}$ for the $I\times I$ bulk and $g^2=\tfrac{209}{450}\approx0.464$ for the interface theory, not appearing as part of the classification (\ref{eq:surfOBdiag}).}
\label{fig:OC2S}
\end{figure}

We can then turn to the case when both sides of the interface are non-free CFTs. For instance, one could consider an $O(N)$--Hypercubic interface, which one can study with the choice of bulk tensor \eqref{eq:ONCubiccurve}. Again looking for diagonal solutions, the analysis proceed identically to before, with the couplings divided into two equal groups, as in \eqref{eq:surfsplit}. One can solve the resulting beta function for generic $\alpha$, but the $\alpha$-dependence of the resulting solutions is quite complicated, and we thus just list the fixed points for the RG interface where $\alpha=1/2$,
\begin{equation}\label{eq:surfOBdiag}
    \mathfrak{h}=\frac{N (9 N+44)-8 (N+2) p+16+\sqrt{Q_N}}{12N(N+8)}\,,\qquad \mathfrak{g}=\frac{N (N+28)+8 (N+2) p+16-\sqrt{Q_N}}{12N(N+8)}\,,
\end{equation}
where $Q_N=(N (N+28)+16)^2+64 (N+2)^2 p^2-64 N (N+2)^2 p$. This situation for $N=2$ is shown in Figure \ref{fig:OC2S}, where we have also included the extra non-diagonal fixed point in both the Cubic, and interface setups. One sees that as the symmetry breaking is the same everywhere, the number of solution families matches the number of solutions in the purely Cubic bulk.

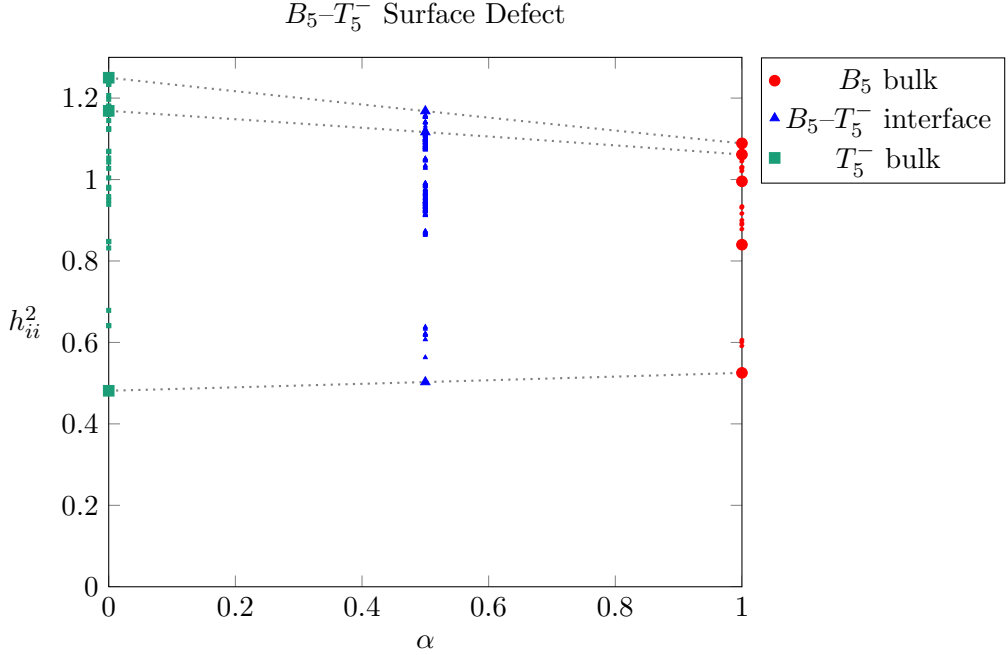
\begin{figure}[H]
\centering
\begin{tikzpicture}
\begin{axis}[
    xmin=-0, xmax=1,
    ymin=0, ymax=1.3,
    xlabel= $\alpha$,
    ylabel= $h_{ii}^2$,
    ylabel style={rotate=-90},
    title={$B_5$--$T_5^-$ Surface Defect},
    legend pos=outer north east,
    reverse legend
]
\addplot[gray,dotted,thick,
domain=0:1,samples=25]
{(-15+x)*(-33-x-sqrt((9-3*x)*(21+x)))/600};
\addplot[gray,dotted,thick,
domain=0:1,samples=25]
{(-15+x)*(-33-x+sqrt((9-3*x)*(21+x)))/600};
\addplot[gray,dotted,thick,
domain=0:1,samples=25]
{(-15+x)^2/180};
\addplot+[
    only marks,
    mark=triangle*,
    mark options={color=blue},
    mark size=0.75pt]
table{Data/tetcub5nondiag.dat};
\addplot+[
    only marks,
    mark=square*,
    mark options={color=Dark2-A},
    mark size=0.75pt]
table{Data/tet5nondiag.dat};
\addplot+[
    only marks,
    mark=*,
    mark options={color=red},
    mark size=0.75pt]
table{Data/cub5nondiag.dat};
\addplot[only marks,mark=square*,mark options={color=Dark2-A},mark size=2pt] coordinates {(0,1.25) (0,0.481307) (0,1.16869)};
\addlegendentry{$T_5^-$ bulk}
\addplot[only marks,mark=triangle*,mark options={color=blue},mark size=2pt] coordinates {(0.5, 1.16806) (0.5, 0.502705) (0.5, 1.11646)};
\addlegendentry{$B_5$--$T_5^-$ interface}
\addplot[only marks,mark=*,mark options={color=red},mark size=2pt] coordinates {(1., 0.84) (1., 0.995556) (1., 1.08889) (1., 0.525254) (1., 
1.06141)};
\addlegendentry{$B_5$ bulk}
\end{axis}
\end{tikzpicture}
    \caption{Solutions to the surface defect beta function \eqref{eq:surfbeta}, where $\lambda_{ijkl}$ takes values along the curve \eqref{eq:CTcurve} with $N=5$. The diagonal solutions lie on the families shown as dotted lines, with the physical conformal defects shown. The non-diagonal fixed points are distinguished as the smaller points, though for clarity we do not show the trajectories here.}
\label{fig:CT5S}
\end{figure}

One can then also consider the case of the Hypercubic--Hypertetrahedral setup, again considering the $T_N^-$ fixed point. Again, we can straightforwardly study the diagonal fixed points analytically. The presence of a $\sum_\alpha e^{\alpha}_{i}e^{\alpha}_je^\alpha_k e^\alpha_l h_{kl}$ term will enforce the condition \eqref{eq:surftetconddiag} for all $\alpha\neq 1$, and thus there will only be three families of solutions, which interpolate to the $p=1$, $p=N-1$ and $p=N$ diagonal solutions in the hypercubic bulk. Again, the $\alpha$-dependence of the generic solutions is quite complicated, and we list only the fixed points for the RG interface,
\begin{equation}
\begin{aligned}
    \mathfrak{h}&=h_{NN}=\frac{(N^2+6N+3)}{3N(N+3)}\varepsilon\,, &&{}\\
    \mathfrak{h}&=\frac{N^2+8N+7-\sqrt{ R_N}}{6
   N (N+3)}\varepsilon\,, &\qquad h_{NN} &=\frac{3N^2+8N-1+\sqrt{R_N}}{6
   N (N+3)}\varepsilon\,,\\
    \mathfrak{h}&=\frac{N^2+8N+7+\sqrt{R_N}}{6
   N (N+3)}\varepsilon\,,\ & h_{NN}&=\frac{3N^2+8N-1-\sqrt{R_N}}{6
   N (N+3)}\varepsilon\,,\\
\end{aligned}
\end{equation}
where $R_N=N(N+2)(N^2+6N+18)+16$. This is shown for $N=5$ in Figure \ref{fig:CT5S}, where one sees that the fixed point in the RG interface simply interpolates between the two points in either bulk. For the non-diagonal fixed points, we must work numerically, sweeping through the space of theories by selecting a number of random starting values for $h_{ij}$ and then using \textit{Mathematica}'s \texttt{FindRoot} function to determine the nearest fixed point. As with the line defect in the Hypercubic--Hypertetrahedral interface, we find that the significantly decreased global symmetry of the system leads to an explosion in the number of solutions for $0<\alpha<1$, and thus namely a much larger number of non-diagonal fixed points in the RG interface. Beginning with $10,000$ starting points, we find 29 non-diagonal fixed points in the $T_5^-$ model, 21 in the $B_5$ model, and 392 in the $B_5$--$T_5^-$ interface. These points are also displayed in Figure \ref{fig:CT5S}, though we do not display the complicated structure of families these non-diagonal points lie on.

\section{CFT Data on the Cubic Interface}
\label{sec:data}

In this section we compute some CFT data in the presence of cubic interactions on the interface. Indeed, the presence of the interface will break the original conformal invariance $SO(d+1,1)$ to the subgroup $SO(d,1)$. This will lead to new CFT data such as non-zero one-point functions that were zero without the interface. 
In the first subsection, we will compute the one-point functions of $\phi^4$ and $\phi^2$. In the next subsection we will compute the two-point function between two $\phi^2$ operators which gets corrections in the presence of the interface. Finally in the last subsection we compute the free energy at leading order. 

\subsection{One-Point Functions}

\begin{figure}[htbp]
    \centering
    \captionsetup[subfigure]{labelformat=empty}
    \subfloat[]{\begin{tikzpicture}[baseline=(vert_cent.base), square/.style={regular polygon,regular polygon sides=4},scale=0.4]
        \draw[densely dashed] (0,3)--(0,-3);
        \draw[very thick] (0,0) circle (2cm);
        \node[fill=white,inner sep=-0.5pt,scale=1.27] at (-2,0) (l) {$\otimes$};
        \node at (2,0) [square, draw,fill=white,inner sep=1.2pt,outer sep=0pt]  (r) {};
        \draw[very thick] (l) to[out=-30, in=-150] (r);
        \draw[very thick] (l) to[out=30, in=150] (r);
        \node (vert_cent) at (current bounding box.center) {};
    \end{tikzpicture}}
    \hspace{2cm}
    \subfloat[]{\begin{tikzpicture}[baseline=(vert_cent.base), square/.style={regular polygon,regular polygon sides=4},scale=0.4]
        \draw[densely dashed] (0,3)--(0,-3);
        \draw[very thick] (-3,0) circle (2cm);
        \node[fill=white,inner sep=-0.5pt,scale=1.27] at (-5,0) (l) {$\otimes$};
        \node at (-1,0) [square, draw,fill=white,inner sep=1.2pt,outer sep=0pt]  (r) {};
        \draw[very thick] (l) to[out=-30, in=-150] (r);
        \draw[very thick] (l) to[out=30, in=150] (r);
        \node (vert_cent) at (current bounding box.center) {};
    \end{tikzpicture}}
    \hspace{2cm}
    \subfloat[]{\begin{tikzpicture}[baseline=(vert_cent.base), square/.style={regular polygon,regular polygon sides=4},scale=0.4]
        \node at (0,2) [circle, draw,fill=white,inner sep=1.2pt,outer sep=0pt]  (i1) {};
        \node at (0,-2) [circle, draw,fill=white,inner sep=1.2pt,outer sep=0pt]  (i2) {};
        \node[] at (-3,0) (p1) {};
        \node[] at (-1,0) (p2) {};
        \draw[very thick] (i1) to [bend right=20] (p1);
        \draw[very thick] (i1) to [bend left=20] (p1);
        \draw[very thick] (i2) to [bend right=20] (p1);
        \draw[very thick] (i2) to [bend left=20] (p1);
        \draw[very thick] (i1) to [bend left] (i2);
        \draw[densely dashed] (0,3)--(0,-3);
        \node[inner sep=0pt,outer sep=0pt, scale=1.27]  at (-3,0) {$\otimes$};
        \node (vert_cent) at (current bounding box.center) {};
    \end{tikzpicture}}
    \caption{The three diagrams contributing to $\langle\phi_I^4\rangle_{2}$ at $O(\varepsilon)$. The vertical dashed line represent the interface localised at $(\mathbf{x},0)$. The white square vertices represent bulk couplings: $\lambda^1$ on the right and $\lambda^2$ on the left, while the round vertices represent interface couplings. We obtain $\langle\phi_I^4\rangle_{1}$ by inserting the quartic operator on the right of the interface.}
    \label{fig:onepointphi4}
\end{figure}
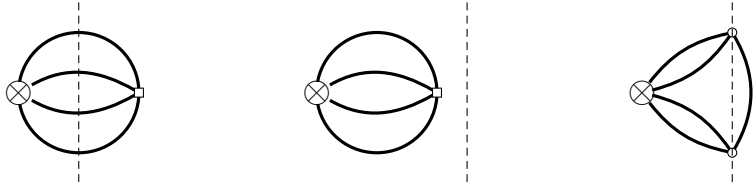

\paragraph{One-point function of $\phi^4_I$} Consider the one-point function of the quartic operator $\phi^4_I$, where $I=(ijkl)$ represents a symmetrised generalised index. The value of this one-point function will depend on which side of the interface we insert the operator, which we can indicate by the use of subscript $1$ or $2$. Mindful that $\lambda^1$ and $\lambda^2$ will include explicit factors of $\varepsilon$, at $O(\varepsilon)$ and $h$ will include factors of $\sqrt{\varepsilon}$ only the diagrams shown in Figure \ref{fig:onepointphi4} will contribute.

The contribution from the graph on the left of Figure \ref{fig:onepointphi4} is given by
\begin{equation}
\begin{split}
&-\lambda^1_{ijkl}C_\phi^4\int d^{d-1}\vec{y}\int_0^\infty d y_\perp \frac{1}{((\vec{y}-\vec{x})^2+(y_\perp-x_\perp)^2)^{2d-4}}=-\frac{1}{512\pi^4|x_\perp|^4}\left(\frac{\lambda^1_{ijkl}}{16\pi^2}\right)+O(\varepsilon^2)\, ,
\end{split}
\end{equation}
where the integral over $y\geq0$ has been performed straightforwardly and $C_{\phi}=\frac{\Gamma(\tfrac{d-2}{2})}{4\pi^{\tfrac{d}{2}}}$. Note that in this section we do not rescale any of the couplings by factors of $\pi$.

As this one-point function must vanish if we make $CFT_1$ and $CFT_2$ identical, it must be proportional to $\lambda^1$ and $\lambda^2$ only in the combination $\lambda^1_{ijkl}-\lambda^2_{ijkl}$, so that the contribution from the graph in the middle of Figure \ref{fig:onepointphi4} is
\begin{equation}
    \frac{1}{512\pi^4|x_\perp|^4}\left(\frac{\lambda^2_{ijkl}}{16\pi^2}\right)+O(\varepsilon^2)\,.
\end{equation}

The contribution from the insertion of defect couplings is given by:
\begin{align}
    &C_{\phi}^5(h_{ijm}h_{klm} + \text{Perms.})\int d^{d-1}  \mathbf{z}_1 \int d^{d-1}  \mathbf{z}_2 \frac{1}{( \mathbf{z}_1^2+x_{\perp}^2)^{d-2}( \mathbf{z}_2^2+x_{\perp}^2)^{d-2}(( \mathbf{z}_1- \mathbf{z}_2)^2)^{\tfrac{d-2}{2}}} \crcr
    =&\frac{C_{\phi}^5(h_{ijm}h_{klm}+ \text{Perms.})}{\Gamma(d-2)^2\Gamma(\tfrac{d}{2}-1)}\int_0^{\infty}da\,db\,dc \frac{(ab)^{d-3}c^{\tfrac{d}{2}-2}}{((a+b)c+ab)^{\tfrac{d-1}{2}}}e^{-x_{\perp}^2(a+b)} \crcr
    =&C_{\phi}^5(h_{ijm}h_{klm} + \text{Perms.})\frac{\pi^4}{4|x_\perp|^4}=\frac{(h_{ijm}h_{klm} + \text{Perms.})}{(4\pi)^6 |x_\perp|^4} \, ,
    \label{eq:onepointphi4h2}
\end{align}
where we set $d=4$ and used the following formulas to integrate the Schwinger parameters:
\begin{equation}
    \int_0^{\infty}dc \frac{c^{u-1}}{(c+\gamma)^v}=\gamma^{u-v}\frac{\Gamma(u)\Gamma(v-u)}{\Gamma(v)} \quad ,   \, \text{Re}(u)\geq 0 \, , \, \text{Re}(v)\geq \text{Re}(u ) \, ,
    \label{eq:help1}
\end{equation}
and
\begin{equation}
    \int_0^{\infty} da db \frac{(a b)^{u-1}}{(a+b)^{v}}e^{-(a+b)}=\frac{\Gamma(u)^2\Gamma(2u-v)}{\Gamma(2u)} \quad ,   \, \text{Re}(u)\geq 0 \, , \, 2\text{Re}(u)\geq \text{Re}(v) \, .
    \label{eq:help2}
\end{equation}

Putting all the contributions together we obtain
\begin{align}
    \langle\phi_I^4(\vec{x},x_\perp)\rangle_{2}&=\frac{1}{(4\pi)^6|x_\perp|^4}\left(\frac{\lambda^2_{ijkl}-\lambda^1_{ijkl}}{2}+ (h_{ijm}h_{klm} + \text{Perms.})\right) \, .
\end{align}

Similarly, we can evaluate the diagrams when the operator is inserted on the right, with the integral over $y\leq0$ being trivial. We obtain
\begin{align}
    \langle\phi_I^4(\vec{x},x_\perp)\rangle_{1}&=\frac{1}{(4\pi)^6|x_\perp|^4}\left(\frac{\lambda^1_{ijkl}-\lambda^2_{ijkl}}{2}+ (h_{ijm}h_{klm} + \text{Perms.})\right) \, .
\end{align}

One notices immediately that the one-point function coefficients are equal and opposite for operators inserted on opposite sides of the interface. These results match those found in \cite{Gliozzi:2015qsa}\footnote{One must be mindful that they have defined their fields $\phi_i$ with a factor of $2\pi$ relative to ours.} once one sets
\begin{equation}
    \lambda^2_{ijkl}=0\,,\qquad\qquad \lambda^1_{ijkl}=\frac{\varepsilon}{N+8}\big(\delta_{ij}\delta_{kl}+\text{Perms.}\big)\,,
\end{equation}
and contracts with the appropriate external tensor to set the operator insertion to $(\phi^2)^2$\,.

\begin{figure}[htbp]
    \centering
    \begin{tikzpicture}[baseline=(vert_cent.base), square/.style={regular polygon,regular polygon sides=4},scale=0.4]
        \node at (0,2) [circle, draw,fill=white,inner sep=1.2pt,outer sep=0pt]  (i1) {};
        \node at (0,-2) [circle, draw,fill=white,inner sep=1.2pt,outer sep=0pt]  (i2) {};
        \node[] at (-3,0) (p1) {};
        \node[] at (-1,0) (p2) {};
        \draw[very thick] (i1) -- (p1);
        \draw[very thick] (i2)-- (p1);
        \draw[very thick] (i1) to [bend left] (i2);
        \draw[very thick] (i1) to [bend right] (i2);
        \draw[densely dashed] (0,3)--(0,-3);
        \node[inner sep=0pt,outer sep=0pt, scale=1.27]  at (-3,0) {$\otimes$};
        \node (vert_cent) at (current bounding box.center) {};
    \end{tikzpicture}
    \caption{Diagram contributing to the one-point function of $\phi^2_{ij}$ at lowest order. The vertical dashed line represent the interface localised at $(\mathbf{x},0)$. The round vertices represent interface couplings.}
    \label{fig:onepointphi2}
\end{figure}
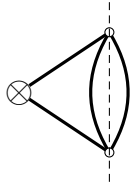

\paragraph{One-point function of $\phi^2_{ij}$} Let us now compute the one-point function of the quadratic operator $\phi^2_{ij}$. At lowest order, we only have one contribution given by the diagram of Figure \ref{fig:onepointphi2}. Since this diagram involves only interface couplings we will obtain the same result wether we insert the operator on the right or on the left of the interface. 

We thus have
\begin{align}
    \langle\phi_{ij}^2(\vec{x},x_\perp)\rangle&=\frac{1}{2}h_{ikl}h_{jkl}C_{\phi}^4 \int d^{d-1}  \mathbf{z}_1 \int d^{d-1}  \mathbf{z}_2 \frac{1}{( \mathbf{z}_1^2+x_{\perp}^2)^{\tfrac{d-2}{2}}( \mathbf{z}_2^2+x_{\perp}^2)^{\tfrac{d-2}{2}}(( \mathbf{z}_1- \mathbf{z}_2)^2)^{d-2}} \crcr
    &=\frac{1}{2}h_{ikl}h_{jkl}C_{\phi}^4\frac{\pi^4}{2|x_\perp|^2}=\frac{h_{ikl}h_{jkl}}{2(4\pi)^4|x_\perp|^2} \, .
\end{align}

\subsection{Two-Point Function \texorpdfstring{$\langle \phi_{ij}^2 \phi_{kl}^2 \rangle$}{phi phi}}

\begin{figure}[htbp]
    \centering
    \captionsetup[subfigure]{labelformat=empty}
    \subfloat[]{\begin{tikzpicture}[baseline=(vert_cent.base), square/.style={regular polygon,regular polygon sides=4},scale=0.6]
        \node at (2,3) [square, draw,fill=white,inner sep=1.2pt,outer sep=0pt]  (r) {};
        \node[] at (-3,0) (p1) {};
        \node[] at (-1,0) (p2) {};
        \draw[very thick] (r) to [bend right] (p1);
        \draw[very thick] (r) to [out=185,in=40] (p1);
        \draw[very thick] (r) to [out=-130,in=35] (p2);
        \draw[very thick] (r) to [bend left] (p2);
        \draw[densely dashed] (0,4)--(0,-2);
        \draw[densely dashed] (-5,0)--(4,0);
        \node[inner sep=0pt,outer sep=0pt, scale=1.27]  at (-3,0) {$\otimes$};
        \node[inner sep=0pt,outer sep=0pt, scale=1.27]  at (-1,0) {$\otimes$};
        \node (vert_cent) at (current bounding box.center) {};
    \end{tikzpicture}
}
\hspace{1cm}
\subfloat[]{\begin{tikzpicture}[baseline=(vert_cent.base), square/.style={regular polygon,regular polygon sides=4},scale=0.6]
        \node at (-2,3) [square, draw,fill=white,inner sep=1.2pt,outer sep=0pt]  (r) {};
        \node[] at (-3,0) (p1) {};
        \node[] at (-1,0) (p2) {};
        \draw[very thick] (r) to [bend right] (p1);
        \draw[very thick] (r) to [out=-100,in=45] (p1);
        \draw[very thick] (r) to [out=-80,in=135] (p2);
        \draw[very thick] (r) to [bend left] (p2);
        \draw[densely dashed] (0,4)--(0,-2);
        \draw[densely dashed] (-5,0)--(4,0);
        \node[inner sep=0pt,outer sep=0pt, scale=1.27]  at (-3,0) {$\otimes$};
        \node[inner sep=0pt,outer sep=0pt, scale=1.27]  at (-1,0) {$\otimes$};
        \node (vert_cent) at (current bounding box.center) {};
    \end{tikzpicture}
} \\
\subfloat[]{\begin{tikzpicture}[baseline=(vert_cent.base), square/.style={regular polygon,regular polygon sides=4},scale=0.6]
        \node at (0,3) [circle, draw,fill=white,inner sep=1.2pt,outer sep=0pt]  (i1) {};
        \node at (0,1) [circle, draw,fill=white,inner sep=1.2pt,outer sep=0pt]  (i2) {};
        \node[] at (-3,0) (p1) {};
        \node[] at (-1,0) (p2) {};
        \draw[very thick] (i1) to [bend right] (p1);
        \draw[very thick] (i1) to [bend left] (p1);
        \draw[very thick] (i2) to [bend right] (p2);
        \draw[very thick] (i2) to [bend left] (p2);
        \draw[very thick] (i1) to [bend left] (i2);
        \draw[densely dashed] (0,4)--(0,-2);
        \draw[densely dashed] (-5,0)--(3,0);
        \node[inner sep=0pt,outer sep=0pt, scale=1.27]  at (-3,0) {$\otimes$};
        \node[inner sep=0pt,outer sep=0pt, scale=1.27]  at (-1,0) {$\otimes$};
        \node (vert_cent) at (current bounding box.center) {};
    \end{tikzpicture}
} \hspace{0.75cm}
\subfloat[]{\begin{tikzpicture}[baseline=(vert_cent.base), square/.style={regular polygon,regular polygon sides=4},scale=0.6]
        \node at (0,3) [circle, draw,fill=white,inner sep=1.2pt,outer sep=0pt]  (i1) {};
        \node at (0,1) [circle, draw,fill=white,inner sep=1.2pt,outer sep=0pt]  (i2) {};
        \node[] at (-3,0) (p1) {};
        \node[] at (-1,0) (p2) {};
        \draw[very thick] (i1) -- (p1);
        \draw[very thick] (i1) -- (p2);
        \draw[very thick] (i2)-- (p2);
        \draw[very thick] (i2)-- (p1);
        \draw[very thick] (i1) to [bend left] (i2);
        \draw[densely dashed] (0,4)--(0,-2);
        \draw[densely dashed] (-5,0)--(3,0);
        \node[inner sep=0pt,outer sep=0pt, scale=1.27]  at (-3,0) {$\otimes$};
        \node[inner sep=0pt,outer sep=0pt, scale=1.27]  at (-1,0) {$\otimes$};
        \node (vert_cent) at (current bounding box.center) {};
    \end{tikzpicture}
} 
\hspace{0.75cm}
\subfloat[]{\begin{tikzpicture}[baseline=(vert_cent.base), square/.style={regular polygon,regular polygon sides=4},scale=0.6]
        \node at (0,3) [circle, draw,fill=white,inner sep=1.2pt,outer sep=0pt]  (i1) {};
        \node at (0,1) [circle, draw,fill=white,inner sep=1.2pt,outer sep=0pt]  (i2) {};
        \node[] at (-3,0) (p1) {};
        \node[] at (-1,0) (p2) {};
        \draw[very thick] (i1) -- (p1);
        \draw[very thick] (i2)-- (p2);
        \draw[very thick] (p2) to [bend left] (p1);
        \draw[very thick] (i1) to [bend left] (i2);
        \draw[very thick] (i1) to [bend right] (i2);
        \draw[densely dashed] (0,4)--(0,-2);
        \draw[densely dashed] (-5,0)--(3,0);
        \node[inner sep=0pt,outer sep=0pt, scale=1.27]  at (-3,0) {$\otimes$};
        \node[inner sep=0pt,outer sep=0pt, scale=1.27]  at (-1,0) {$\otimes$};
        \node (vert_cent) at (current bounding box.center) {};
    \end{tikzpicture}
} 
    \caption{Contributions to the two-point function $\langle \phi^2(-y_1) \phi^2(-y_2) \rangle$ in the colinear geometry. The vertical dashed line represent the interface localised at $(\mathbf{x},0)$ and the horizontal dashed line represents the line $(\mathbf{0},y)$. }
    \label{fig:twopointleft}
\end{figure}
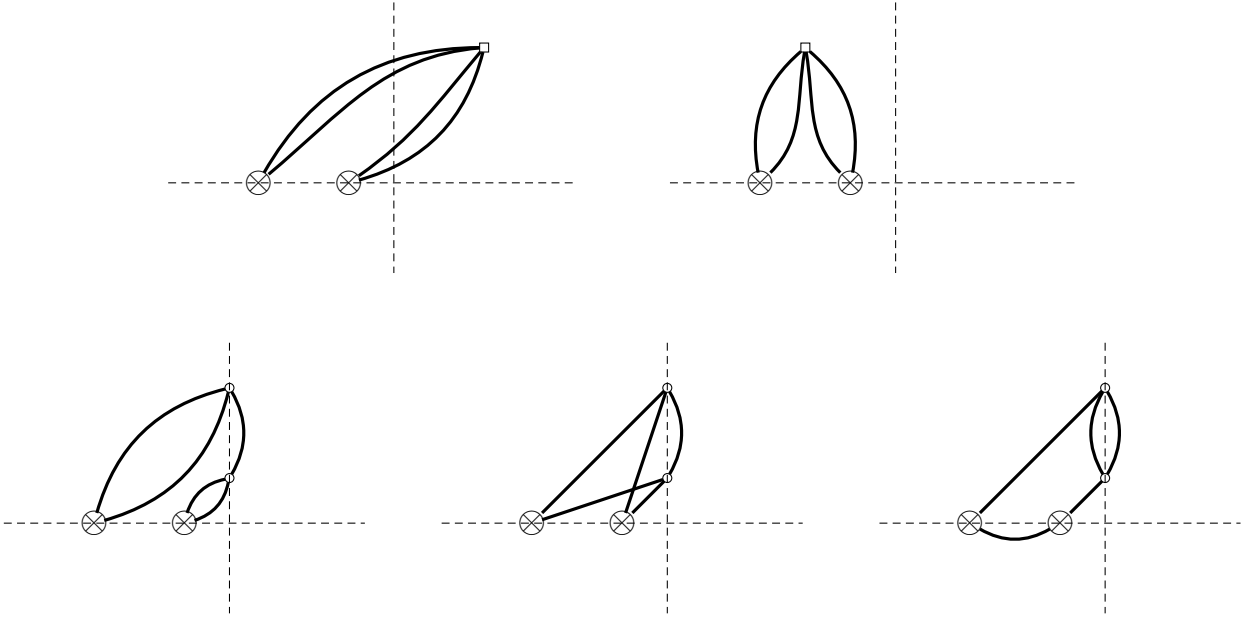

In this subsection we compute the two-point function $\langle \phi_{ij}^2 \phi_{kl}^2 \rangle$ on the left side of the interface. At one loop there are five contributions: we can insert a bulk coupling on the left of the interface or on the right of the interface or we can insert two defect couplings on the interface in three different ways. As the result will only depend on a cross-ratio we can do the computation in the colinear geometry of Figure \ref{fig:twopointleft} and compute $\langle \phi^2(-x_{\perp}) \phi^2(-y_{\perp}) \rangle$ with $x_{\perp},y_{\perp}>0$. 

The contribution from an insertion of a bulk coupling on the right of the interface was computed in \cite{Gliozzi:2015qsa} and is given by:
\begin{align}
    &-\lambda_{ijkl}^{1} C_\phi^4\int_0^{\infty} dz_{\perp} \int d^{d-1} \mathbf{z} \frac{1}{\left(   \mathbf{z}^2 +(z_{\perp}+x_{\perp})^2 \right)^{d-2}\left(   \mathbf{z}^2 +(z_{\perp}+y_{\perp})^2 \right)^{d-2}}\crcr
    =&-\frac{ \lambda_{ijkl}^{1}}{256\pi^6(x_{\perp}-y_{\perp})^4}\left(\frac{\xi}{\xi+1} -\ln(\xi +1)\right) \, ,
    \label{eq:Gliozzi}
\end{align}
where we set $d=4$ and $\xi$ is the cross-ratio
\begin{equation}
    \xi=\frac{(x_{\perp}-y_{\perp})^2}{4x_{\perp} y_{\perp}} \, .
\end{equation}

The insertion of a bulk coupling on the left of the interface is given by
\begin{align}
\begin{aligned}
    & - \lambda^2_{ijkl} C_\phi^4 \int d^{d-1} \mathbf{z}\int_{-\infty}^0 d z_\perp \frac{1}{\left( \mathbf{z}^2+(z_\perp + x_\perp)^2\right)^{d-2}\left( \mathbf{z}^2+(z_\perp + y_\perp)^2\right)^{d-2}} \\
    = & -\lambda^2_{ijkl} C_\phi^4 \int d^{d-1} \mathbf{z}\int d z_\perp \frac{1}{\left( \mathbf{z}^2+(z_\perp + x_\perp)^2\right)^{d-2}\left( \mathbf{z}^2+(z_\perp + y_\perp)^2\right)^{d-2}} \\
    {} & + \lambda^2_{ijkl} C_\phi^4 \int d^{d-1} \mathbf{z}\int_0^\infty d z_\perp \frac{1}{\left( \mathbf{z}^2+(z_\perp + x_\perp)^2\right)^{d-2}\left( \mathbf{z}^2+(z_\perp + y_\perp)^2\right)^{d-2}} 
    , .
\end{aligned}
\end{align}
Note that the second term is the same as the right hand contribution, but with the coupling of $\text{CFT}_2$. The first term here can be simplified as
\begin{align}
    - \lambda^2_{ijkl} C_\phi^4 \int d^d z \frac{1}{((z+x)^2)^{d-2}((z+y)^2)^{d-2}},
    \label{eq:WPnote1}
\end{align}
where $z=( \mathbf{z},z_{\perp}), x=(\mathbf{0},-x_{\perp}), y=(\mathbf{0},-y_{\perp})$. It can be calculated perturbatively around $d=4-\varepsilon$ to be
\begin{align}
    - \lambda^2_{ijkl} C_\phi^4\frac{\Gamma\left(\frac{3d-8}{2}\right)\Gamma\left(\frac{4-d}{2}\right)^2}{\Gamma(d-2)^2\Gamma(4-d)}\frac{\pi^{d/2}}{\left((x-y)^2\right)^{\frac{3d-8}{2}}} = -\lambda_{ijkl}^2\frac{1}{4\pi^4|x_{\perp}-y_{\perp}|^4}\left(\frac{1}{16\pi^2\varepsilon}\right) + \mathcal{O}(\varepsilon)\,.
\label{eq:divtwopoint}
\end{align}

The operator $\phi^2$ which appears in this correlator will be renormalised by the $\lambda^2_{ijkl}$ interaction present to the left of the interface. As this renormalisation is associated with local divergences, it is unaffected by the presence of the interface, and for the purposes of computing it we can treat $\lambda^2$ as filling the space. We can compute the renormalisation of $\phi^2$ using the one-loop diagram for the correlation function $\langle\widetilde{\phi^2_{ij}}(p_1-p_2)\widetilde{\phi_k}(-p_1)\widetilde{\phi_l}(p_2)\rangle$
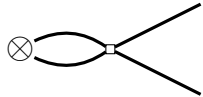
$$\begin{tikzpicture}[baseline=(vert_cent.base), square/.style={regular polygon,regular polygon sides=4},scale=0.6]
        \node at (0,0) [square, draw,fill=white,inner sep=1.2pt,outer sep=0pt]  (i1) {};
        \node[inner sep=0pt,outer sep=0pt, scale=1.27] (p1) at (-2,0) {$\otimes$};
        \draw[very thick] (p1) to [bend left] (i1);
        \draw[very thick] (p1) to [bend right] (i1);
        \draw[very thick] (i1) to (2,1);
        \draw[very thick] (i1) to (2,-1);
        \node (vert_cent) at (current bounding box.center) {};
    \end{tikzpicture}$$
which is given by the integral
\begin{equation}
    -\frac{\lambda^2_{ijkl}}{p_1^2 p_2^2}\int\frac{d^dk}{(2\pi)^d}\frac{1}{(p_1+k)^2(p_2+k)^2}=-\frac{\lambda^2_{ijkl}}{(8\pi^2)p_1^2p_2^2\varepsilon}+O(\varepsilon^0)\,.
\end{equation}
The divergence in this diagram necessitates the introduction of the renormalised operator
\begin{equation}[\phi_{ij}^2]=\phi_{ij}^2+\frac{\lambda^2_{ijkl}}{(16\pi^2)\varepsilon}\phi^2_{kl}\,.
\end{equation}
Using the renormalised operator in the two-point function removes the $1/\varepsilon$ pole from \eqref{eq:divtwopoint}, and replaces it with the finite result
\begin{align}
    \frac{\lambda^2_{ijkl}}{16\pi^2}\frac{-1+2\log\big(\tfrac{2}{\mu|x_\perp-y_\perp|}\big)}{8\pi^4|x_\perp-y_\perp|^4}+O(\varepsilon)\,,
\end{align}
where we have rescaled the RG parameter to $\mu^2\rightarrow\mu^2e^{-\frac{\gamma_E}{2}}/4\pi$.

The first contribution from an insertion of two defect couplings on the interface is given by

\begin{align}
 &h_{ijm}h_{klm} C_\phi^5\int d^{d-1}  \mathbf{z}_1 \int d^{d-1}  \mathbf{z}_2\frac{1}{( \mathbf{z}_1^2+x_{\perp}^2)^{d-2}( \mathbf{z}_2^2+y_{\perp}^2)^{d-2}(( \mathbf{z}_1- \mathbf{z}_2)^2)^{\tfrac{d-2}{2}}} \crcr 
 & = \frac{4\pi^{d-\tfrac{3}{2}}h_{ijm}h_{klm}}{\Gamma(\tfrac{d-1}{2})\Gamma(\tfrac{d-2}{2})} C_\phi^5\int_0^{\infty} dz_1\int_0^{\infty} dz_2 \int_0^{\pi} d\theta\frac{z_1^{d-2}z_2^{d-2}\sin(\theta)^{d-3}}{(z_1^2+x_{\perp}^2)^{d-2}(z_2^2+y_{\perp}^2)^{d-2}(z_1^2+z_2^2+2z_1z_2\cos(\theta))^{\tfrac{d-2}{2}}}\crcr
 &= \frac{\pi^4 h_{ijm}h_{klm}}{x_{\perp} y_{\perp}(x_{\perp}+y_{\perp})^2} C_\phi^5 = h_{ijm}h_{klm} \frac{1}{256\pi^6(x_\perp-y_\perp)^4}\frac{\xi^2}{\xi+1}\, ,
 \label{eq:2ptdefectcont1}
\end{align}
where we again set $d=4$ after changing variable to spherical coordinates.
Notice that if we take the limit $y_{\perp} \rightarrow x_{\perp}$, we recover the one-point function given in \eqref{eq:onepointphi4h2}.

Similarly, in $d=4$, the second contribution from an insertion of two defect couplings on the interface is given by 
\begin{align}
 &h_{ikm}h_{jlm} C_\phi^5\int d^{d-1} \mathbf{z}_1 \int d^{d-1} \mathbf{z}_2 \frac{1}{(\mathbf{z}_1^2+x_{\perp}^2)^{\tfrac{d-2}{2}}(\mathbf{z}_1^2+y_{\perp}^2)^{\tfrac{d-2}{2}}(\mathbf{z}_2^2+x_{\perp}^2)^{\tfrac{d-2}{2}}(\mathbf{z}_2^2+y_{\perp}^2)^{\tfrac{d-2}{2}}((\mathbf{z}_1-\mathbf{z}_2)^2)^{\tfrac{d-2}{2}}} \crcr 
 & = \frac{4\pi^4 h_{ikm}h_{jlm} \ln(\xi+1)}{(x_{\perp}^2-y_{\perp}^2)^2}  C_\phi^5  = \frac{ h_{ikm}h_{jlm} \ln(\xi+1)}{256 \pi^6(x_{\perp}- y_{\perp})^4} \frac{\xi}{\xi+1}\, . 
 \label{eq:2ptdefectcont2}
\end{align}
Notice again that, as for the previous term, if we take the limit $y_{\perp} \rightarrow x_{\perp}$, we recover the one-point function given in \eqref{eq:onepointphi4h2}.

The final contribution from an insertion of two defect couplings on the interface is given by 
\begin{align}
 & \frac{1}{2} C_{\phi}^5\delta_{ik}h_{jmn}h_{lmn}\int d^{d-1} \mathbf{z}_1 \int d^{d-1} \mathbf{z}_2 \frac{1}{(\mathbf{z}_1^2+x_{\perp}^2)^{\tfrac{d-2}{2}}(\mathbf{z}_2^2+y_{\perp}^2)^{\tfrac{d-2}{2}}((x_{\perp}-y_{\perp})^2)^{\tfrac{d-2}{2}}((\mathbf{z}_1-\mathbf{z}_2)^2)^{d-2}} \, .
\end{align}

To compute this integral we used partial Fourier transform along the interface directions. After integrating two of the momenta and remembering that a factor of $C_\phi^4$ is included in the partial Fourier transform, we obtain :
\begin{align}
    &\frac{1}{2} C_{\phi}\delta_{ik}h_{jmn}h_{lmn}\int \frac{d^{d-1}\mathbf{p}}{(2\pi)^{d-1}}\int \frac{d^{d-1}\mathbf{q}}{(2\pi)^{d-1}} \frac{e^r{-|x_\perp||\mathbf{p}|}}{2|\mathbf{p}|} \frac{e^{-|y_\perp||\mathbf{p}|}}{2|\mathbf{p}|} \frac{1}{2|\mathbf{p}-\mathbf{q}|} \frac{1}{2|\mathbf{q}|} \crcr
    & = \frac{1}{2} C_{\phi}\delta_{ik}h_{jmn}h_{lmn} \frac{1}{2(4\pi)^{d-1}\pi^{3/2}} \frac{\Gamma\left(\frac{3-d}{2}\right)\Gamma\left(\frac{d-2}{2}\right)}{\Gamma(d-2)} \int_0^\infty dp\, p^{2d-7}e^{-(|x_\perp|+|y_\perp|)p} \, ,
\end{align}
where the $\mathbf{q}$ integral was done using the formula
\begin{equation}
    \int \frac{d^d \mathbf{q}}{(2\pi)^d}\frac{1}{\mathbf{q}^{2\alpha}(\mathbf{p}+\mathbf{q})^{2\beta}}=\frac{|\mathbf{p}|^{d-2\alpha-2\beta}}{(4\pi)^{d/2}}\frac{\Gamma(\tfrac{d}{2}-\alpha)\Gamma(\tfrac{d}{2}-\beta)\Gamma(\alpha+\beta-\tfrac{d}{2})}{\Gamma(\alpha)\Gamma(\beta)\Gamma(d-\alpha-\beta)} \, ,
    \label{eq:bubble}
\end{equation}
and the $\mathbf{p}$ integral was reduce to a one dimensional integral with a change to spherical coordinates. 

The last integral can then be evaluated through integration by parts and we obtain
\begin{align}
    -\frac{1}{2}\delta_{ik}h_{jmn}h_{lmn} C_\phi\frac{1}{(4\pi)^4} \frac{1}{(x_\perp+y_\perp)^2}\frac{1}{(x_\perp-y_\perp)^2} = - \frac{1}{512\pi^4|x_\perp-y_\perp|^4}  \left(\frac{\delta_{ik}h_{jmn}h_{lmn}}{4\pi^2}\right) \frac{\xi}{\xi + 1}.
\end{align}

Gathering all contributions together we finally obtain the two-point function:
\begin{align}
  \langle [\phi^2_{ij}](-x_{\perp}) [\phi^2_{kl}](-y_{\perp}) \rangle=& \bigg(\frac{\delta_{ik}\delta_{jl}}{16\pi^4(x_\perp-y_\perp)^4}+\text{Perms.}\bigg)+\lambda^2_{ijkl}\frac{-1+2\log\big(\tfrac{2}{\mu|x_\perp-y_\perp|}\big)}{128\pi^6(x_\perp-y_\perp)^4}\crcr &-\frac{\lambda_{ijkl}^1-\lambda_{ijkl}^2}{256\pi^6(x_{\perp}-y_{\perp})^4}\left(\frac{\xi}{\xi+1}-\ln(\xi +1)\right) +\frac{h_{ijm}h_{klm}}{256 \pi^6(x_{\perp}-y_{\perp})^4}\frac{\xi^2}{\xi+1}  \crcr 
  &+\bigg(\frac{h_{ikm}h_{jlm}}{256 \pi^6(x_{\perp}-y_{\perp})^4}\frac{\xi}{\xi+1}\ln(\xi +1) +\text{Perms.}\bigg)\crcr 
  &-\left(\frac{\delta_{jl}h_{imn} h_{kmn}}{2048\pi^6(x_{\perp}-y_{\perp})^4}\frac{\xi}{\xi+1}+\text{Perms.}\right) +O(\varepsilon^2)\, ,
  \label{eq:phi2twoptunren}
\end{align}
where there is a sum over distinct permutations of $i$ and $j$ as well as $k$ and $l$. 
The two-point function $\langle [\phi^2_{ij}](x_{\perp}) [\phi^2_{kl}](y_{\perp}) \rangle $ inserted on the right side of the interface is obtained similarly exchanging $\lambda_{ijkl}^1$ and $\lambda_{ijkl}^2$.

Let us now look not at the two-point function of a general quadratic operator, but instead at $\langle\phi^2\phi^2\rangle$, which can be achieved quite simply by contracting \eqref{eq:phi2twoptunren} with $\delta_{ij}\delta_{kl}$. One can expand this two-point function in conformal blocks in the usual way, so that we expect
\begin{equation}\label{eq:confblockexp}
    \langle\phi^2(-x_\perp)\phi^2(-y_\perp)\rangle=\frac{N_{\phi^2}^2}{((x_\perp-y_\perp)^2)^{\Delta_{\phi^2}}}\bigg(1+\sum_{\mathcal{O}\neq\mathbb{I}}\lambda_{\phi^2\phi^2\mathcal{O}}\, a_\mathcal{O}\lsp f_{\text{bulk}}\big(0,\Delta_\mathcal{O},\xi\big)\bigg)\,,
\end{equation}
where $N_{\phi^2}$ is a normalisation factor.

We could be tempted to use the conformal blocks of \cite{Gliozzi:2015qsa} to compute this two-point function. However if we were to do that we would not obtain the same result as in \eqref{eq:phi2twoptunren}. This is because the conformal blocks of \cite{Gliozzi:2015qsa} were obtained using the folding trick from the boundary conformal blocks of \cite{McAvity:1995zd}. Since we have interactions on the interface, they are not valid in our case and the folding trick would have to be re-done carefully to take the interface interactions into account. Equivalently, one could attempt to apply \eqref{eq:confblockexp} directly in the unfolded theory. The conformal blocks would then need to be re-computed for a theory with an interface, which we leave for further work.

\subsection{Free Energy}

In this section we compute the free energy of our system in the presence of the interface at lowest order. In three dimensions, it was shown that the free energy for unitary CFTs is monotonic along the RG flow\cite{Myers:2010xs,Jafferis:2011zi,Klebanov:2011gs,Casini:2012ei}. It has also been shown that a generalized version of the $F$-theorem, in which the monotone is the free energy with a dimensional dependent pre-factor, holds perturbatively at low-loop orders\cite{Fei:2015oha,Pannell:2025ixz}. It is not known if this $F$-theorem holds in the presence of an interface. In the case of a boundary, it was shown that the boundary entropy or $g$-function in two dimensions decreases along RG flows localised on the defect \cite{Affleck:1991tk,Friedan:2003yc,Casini:2016fgb}. Some proposals were made for possible monotones for boundary CFTs in higher dimensions in \cite{Yamaguchi:2002pa,Fujita:2011fp,Nozaki:2012qd,Gaiotto:2014gha,Estes:2014hka,Jensen:2015swa,Herzog:2017kkj,Kobayashi:2018lil,Casini:2018nym,Giombi:2020rmc}. In three dimensions \cite{Jensen:2015swa} showed that the $b$-anomaly coefficient, coefficient of the Euler density term in the boundary trace anomaly, decreases under the boundary RG flow. In \cite{Giombi:2024qbm} it was shown that this $b$-theorem still holds in the case of an interface between a free theory and a $O(N)$ model. However, for a generic interface between two interacting CFTs, no quantity has been proven to decrease under RG flows. Since boundary monotones such as the $g$-function or the $b$-anomaly coefficient can be obtained from the free energy, it is useful to compute the interface free energy in our case in order to explore possible monotones.

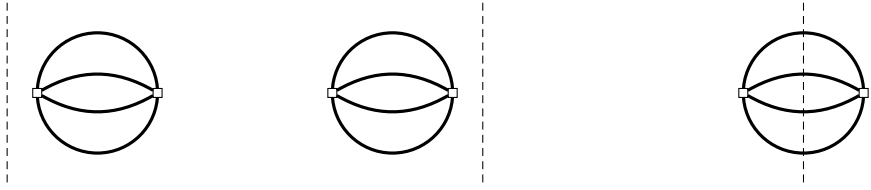
\begin{figure}[H]
    \centering
    \captionsetup[subfigure]{labelformat=empty}
    \subfloat[]{\begin{tikzpicture}[baseline=(vert_cent.base), square/.style={regular polygon,regular polygon sides=4},scale=0.4]
        \draw[densely dashed] (0,3)--(0,-3);
        \draw[very thick] (3,0) circle (2cm);
        \node[] at (-2,0) (l) {};
        \node at (1,0) [square, draw,fill=white,inner sep=1.2pt,outer sep=0pt]  (l) {};
        \node at (5,0) [square, draw,fill=white,inner sep=1.2pt,outer sep=0pt]  (r) {};
        \draw[very thick] (l) to[out=-30, in=-150] (r);
        \draw[very thick] (l) to[out=30, in=150] (r);
        \node (vert_cent) at (current bounding box.center) {};
    \end{tikzpicture}}
    \hspace{2cm}
    \subfloat[]{\begin{tikzpicture}[baseline=(vert_cent.base), square/.style={regular polygon,regular polygon sides=4},scale=0.4]
        \draw[densely dashed] (0,3)--(0,-3);
        \draw[very thick] (-3,0) circle (2cm);
        \node[] at (-5,0) (l) {};
        \node at (-1,0) [square, draw,fill=white,inner sep=1.2pt,outer sep=0pt]  (r) {};
        \node at (-5,0) [square, draw,fill=white,inner sep=1.2pt,outer sep=0pt]  (l) {};
        \draw[very thick] (l) to[out=-30, in=-150] (r);
        \draw[very thick] (l) to[out=30, in=150] (r);
        \node (vert_cent) at (current bounding box.center) {};
    \end{tikzpicture}}
    \hspace{2cm}
    \subfloat[]{\begin{tikzpicture}[baseline=(vert_cent.base), square/.style={regular polygon,regular polygon sides=4},scale=0.4]
       \draw[densely dashed] (0,3)--(0,-3);
       \draw[very thick] (0,0) circle (2cm);
        \node[] at (-5,0) (l) {};
        \node at (2,0) [square, draw,fill=white,inner sep=1.2pt,outer sep=0pt]  (r) {};
        \node at (-2,0) [square, draw,fill=white,inner sep=1.2pt,outer sep=0pt]  (l) {};
        \draw[very thick] (l) to[out=-30, in=-150] (r);
        \draw[very thick] (l) to[out=30, in=150] (r);
        \node (vert_cent) at (current bounding box.center) {};
    \end{tikzpicture}}
    \caption{The three diagrams contributing to the free energy at $O(\varepsilon)$. The vertical dashed line represent the interface localised at $(\mathbf{x},0)$. The white square vertices represent bulk couplings: $\lambda^1$ on the right and $\lambda^2$ on the left.}
    \label{fig:freeenergy}
\end{figure}

At lowest order there are three contributions to the interface free energy given by the diagrams of Figure \ref{fig:freeenergy}.

The first contribution on the left is given by:
\begin{equation}
    -\frac{1}{48}\lambda^1_{ijkl}\lambda^1_{ijkl}C_\phi^4\int_{\mathbb{H^d}} d^dx d^dy \frac{(2R)^{2d}}{(1+x^2)^d(1+y^2)^d}\left(\frac{2R|x-y|}{\sqrt{(1+x^2)(1+y^2)}}\right)^{8-4d}\,.
\end{equation}
where both integrals are over a half space. We denote $x=(x_{||}, z), y=(y_{||},w)$, so the expression becomes
\begin{align}
\begin{aligned}
    & \pe  -\frac{1}{48}\lambda^1_{ijkl}\lambda^1_{ijkl}C_\phi^4 
    \begin{aligned}[t]
        & \int_0^\infty dz \int d^{d-1} x_{||} \int_0^\infty dw \int d^{d-1}y_{||} \\
        & \frac{(2R)^{2d}}{(1+x_{||}^2 + z^2)^d(1+y_{||}^2 + w^2)^d}\left(\frac{2R \sqrt{(x_{||} - y_{||})^2 + (z-w)^2}}{\sqrt{(1+x_{||}^2 + z^2) (1+y_{||}^2 + w^2)}}\right)^{8-4d}
    \end{aligned}
    \\
    & =  -\frac{1}{48}\lambda^1_{ijkl}\lambda^1_{ijkl}C_\phi^4 (2R)^{8-2d} \int_0^\infty dz \int d^{d-1} x_{||} \int_0^\infty dw \int d^{d-1}y_{||} \frac{ \left((x_{||} - y_{||})^2 + (z-w)^2\right)^{4-2d}}{(1+x_{||}^2 + z^2)^{4-d} (1+y_{||}^2 + w^2)^{4-d}}.
    \end{aligned}
\end{align}

Considering the integral alone, we introduce a Schwinger parameter for each term in the fraction. We obtain:
\begin{align}
    I=\frac{1}{\Gamma(4-d)^2\Gamma(2d-4)}& \int_0^\infty d\alpha \alpha^{3-d} \int_0^\infty d\beta \beta^{3-d} \int_0^{\infty} d\gamma \gamma^{2d-5}\int_0^\infty dz \int d^{d-1} x_{||} \int_0^\infty dw \int d^{d-1}y_{||} \crcr
    &  e^{-\alpha(1+x_{||}^2 + z^2)} e^{- \beta(1+y_{||}^2 + w^2)}e^{-\gamma\left((x_{||} - y_{||})^2 + (z-w)^2\right)} \crcr
    =\frac{\pi^{d-1}}{\Gamma(4-d)^2\Gamma(2d-4)}&  \int_0^\infty d\alpha \alpha^{3-d} e^{-\alpha} \int_0^\infty d\beta \beta^{3-d} e^{-\beta} \int_0^\infty d\gamma \frac{\gamma^{2d-5} }{\left(\gamma\alpha + \gamma\beta  + \alpha \beta\right)^{\frac{d-1}{2}}} \crcr
        & \int_0^\infty dz e^{-\alpha z^2} \int_0^\infty dw e^{-\beta w^2} e^{-\gamma(z-w)^2},
\end{align}
where we used Gaussian integration to perform the $x_{||}$ and $y_{||}$ integrals.  
The $z,w$ integrals in the last line can be computed with Mathematica, yielding
\begin{align}
\label{eq: intermediate z w integrals to arctan}
    \frac{\pi -  \arctan \left( \frac{\sqrt{\gamma\alpha + \gamma \beta + \alpha \beta}}{\gamma} \right)}{2 \sqrt{\gamma\alpha + \gamma \beta + \alpha \beta}}.
\end{align}

To perform the remaining integrals we will consider the arctan piece separately and use repeatedly \eqref{eq:help1} and \eqref{eq:help2}.
The integral for the $\pi$ term is:
\begin{align}
    I_1&=\frac{\pi^{d}}{2\Gamma(4-d)^2\Gamma(2d-4)}  \int_0^\infty d\alpha \alpha^{3-d} e^{-\alpha} \int_0^\infty d\beta \beta^{3-d} e^{-\beta} \int_0^\infty d\gamma \frac{\gamma^{2d-5} }{(\alpha+\beta)^{\tfrac{d}{2}}\left(\gamma  + \tfrac{\alpha \beta}{\alpha+\beta}\right)^{\frac{d}{2}}} \crcr
    &= \frac{\pi^{d}}{2\Gamma(4-d)^2\Gamma(2d-4)}  \int_0^\infty d\alpha \alpha^{3-d} e^{-\alpha} \int_0^\infty d\beta \beta^{3-d} e^{-\beta} \frac{(\alpha \beta)^{\tfrac{3d-8}{2}}}{(\alpha+\beta)^{2d-4}}\frac{\Gamma(2d-4)\Gamma(\tfrac{8-3d}{2})}{\Gamma(\tfrac{d}{2})} \crcr
    &= \frac{\pi^d\Gamma(\tfrac{d}{2})\Gamma(\tfrac{8-3d}{2})}{2\Gamma(4-d) }= \frac{\pi^4}{36}+\mathcal{O}(\varepsilon)\, ,
\end{align}
where we took $d=4-\varepsilon$ in the last line. 

Computing the arctan piece is complicated. We first rescale $\gamma$ as $\gamma \rightarrow \gamma \frac{\alpha \beta}{\alpha + \beta}$ and then do the change of variable $\alpha=ab \, , \, \beta= a(1-b)$ and perform the integral over $a$. We obtain:

\begin{align}
 I_2&=-\frac{\pi^{d-1}}{2\Gamma(4-d)^2\Gamma(2d-4)}\int_0^\infty d\alpha \int_0^\infty d\beta\,  \frac{(\alpha \beta)^{3-d}}{(\alpha+\beta)^{\tfrac{d}{2}}} e^{-(\alpha+\beta)} \int_0^\infty d\gamma \frac{\gamma^{2d-5} }{\left(\gamma  + \tfrac{\alpha \beta}{\alpha+\beta}\right)^{\frac{d}{2}}} \arctan \left( \frac{\sqrt{\gamma\alpha + \gamma \beta + \alpha \beta}}{\gamma} \right)\crcr
 &=-\frac{\pi^{d-1}}{2\Gamma(4-d)\Gamma(2d-4)}\int_0^1 db \left( b(1-b)\right)^{\tfrac{d-2}{2}}\int_0^{\infty}d\gamma \frac{\gamma^{2d-5}}{(1+\gamma)^{d/2}}\arctan \left( \frac{\sqrt{1+\gamma}}{\gamma \sqrt{b(1-b)}}\right) \crcr
 &=-\frac{\pi^{d-1}}{2\Gamma(4-d)\Gamma(2d-4)}\int_0^{\infty}d\gamma\int_0^1 db \int_0^1 dz \frac{\gamma^{2d-4}}{(1+\gamma)^{\tfrac{d-1}{2}}}\frac{\left( b(1-b)\right)^{\tfrac{d-1}{2}}}{(1+\gamma)z^2 + \gamma^2 b(1-b)} \, ,
\end{align}
where in the last line we have used the integral representation of the arc tangent. 

We will now introduce a Mellin-Barnes parameter to split the second denominator:
\begin{equation}
    \frac{1}{(1+\gamma)z^2 + \gamma^2 b(1-b)}=\int_{\mathcal{C}} \frac{du}{2i\pi}\Gamma(-u)\Gamma(u+1)\frac{(\gamma^2b(1-b))^u}{(z^2(1+\gamma))^{u+1}} \, .
\end{equation}

The integrals over $\gamma$ and $b$ are then Euler Beta functions and the integral over $z$ can be taken as long as the real part of $u$ is less than $-\frac{1}{2}$. Choosing the contour so that all Gamma functions are well-defined we obtain:
\begin{equation}
    I_2=\frac{\pi^{d-1/2}}{2\Gamma(4-d)\Gamma(2d-4)}\int_{(-\tfrac{5}{2}+\tfrac{3\varepsilon}{2})^-}\frac{du}{2i\pi}2^{-4+\varepsilon-2u}\frac{\Gamma(-u)\Gamma(u+1)}{2u+1}\frac{\Gamma(5+2u-2\varepsilon)\Gamma(-\tfrac{5}{2}+\tfrac{3\varepsilon}{2}-u)}{\Gamma(3+u-\tfrac{\varepsilon}{2})} \, .
\end{equation}
The only pole giving a divergence in $1/\varepsilon$ is located at $u=-\tfrac{5}{2}+\tfrac{3\varepsilon}{2}$. We then obtain:

\begin{align}
    I_2&=-\frac{\pi^{d-1/2}}{2\Gamma(4-d)\Gamma(2d-4)}2^{1-2\varepsilon}\frac{\Gamma(5/2-3\varepsilon/2)\Gamma(-3/2+3\varepsilon/2)\Gamma(\varepsilon)}{(4-3\varepsilon)\Gamma(1/2+\varepsilon)}+\mathcal{O}(\varepsilon) \crcr
    &= -\frac{\pi^4}{24} +\mathcal{O}(\varepsilon) \, .
\end{align}
We thus obtain $I=-\frac{\pi^4}{72} +\mathcal{O}(\varepsilon)$. The contribution in the middle of Figure \ref{fig:freeenergy} is the same as the first one with $\lambda_{ijkl}^1$ replaced by $\lambda_{ijkl}^2$. 

The last contribution is given by:
\begin{equation}
    -\frac{1}{24}\lambda^1_{ijkl}\lambda^2_{ijkl}C_\phi^4\int_0^\infty dz \int d^{d-1} x_{||} \int^0_{-\infty} dw \int d^{d-1}y_{||} \frac{(2R)^{2d}}{(1+x^2)^d(1+y^2)^d}\left(\frac{2R|x-y|}{\sqrt{(1+x^2)(1+y^2)}}\right)^{8-4d}\,,
\end{equation}
where $x=(x_{||}, z), y=(y_{||},w)$.
We can compute it following the same steps as for the first term. We get a similar expression but with only the arc tangent piece. 

We finally obtain the free energy at $O(\varepsilon^2)$
\begin{equation}\label{eq:freeenergy}
    F=\frac{1}{3456(4\pi)^4}\left(\lambda^1_{ijkl}\lambda^1_{ijkl}+\lambda^2_{ijkl}\lambda^2_{ijkl}-6\lambda^1_{ijkl}\lambda^2_{ijkl} \right)+O(\varepsilon^3)\, .
\end{equation}
Armed with this expression, we can compute the free energy for the bulk and RG interface models studied earlier in the paper. Notice that setting $\lambda^2_{ijkl}$ to produce a $\text{CFT}_1$--Free interface, produces a free energy which is $-1/4$ that of setting $\lambda^2_{ijkl}=\lambda^1_{ijkl}$. That is,
\begin{equation}
    F_{\text{CFT}_1-\text{Free}}=-\frac{1}{4}F_{\text{CFT}_1}+O(\varepsilon^3)\,.
\end{equation}
Moreover, as $\lambda_{ijkl}\lambda_{ijkl}$ is manifestly positive, we will have $F_{\text{CFT}_1}<0$ for all non-trivial CFTs, and $F_{\text{CFT}_1-\text{Free}}>0$. For example, if we set $\lambda^2_{ijkl}=0$ and $\lambda^1_{ijkl}=(4\pi)^2\frac{\varepsilon}{N+8}\left( \delta_{ij}\delta_{kl} +\delta_{ik}\delta_{jl} + \delta_{il}\delta_{jk}\right)$, we obtain
\begin{equation}
    F=\frac{N(N+2)}{(N+8)^2}\frac{\varepsilon^2}{1152} +\mathcal{O}(\varepsilon^3) \, ,
\end{equation}
which matches the result of \cite{Giombi:2024qbm} where one side of the interface is free and the other side is the $O(N)$ bulk. If we instead set $\lambda^1_{ijkl}=\lambda^2_{ijkl}=(4\pi)^2\frac{\varepsilon}{N+8}\left( \delta_{ij}\delta_{kl} +\delta_{ik}\delta_{jl} + \delta_{il}\delta_{jk}\right)$, we obtain
\begin{equation}
    F_{O(N)}=-\frac{N(N+2)}{(N+8)^2}\frac{\varepsilon^2}{288} + \mathcal{O}(\varepsilon^3)\, ,
\end{equation}
which matches the $O(N)$ free energy without interface computed in \cite{Fei:2015oha}. Setting $\lambda^1_{ijkl}=\lambda^2_{ijkl}=(4\pi)^2\lambda^{B_N}_{ijkl}$ from \eqref{eq:lambdaBN}, we obtain 
\begin{equation}
    F_{B_N}=-\frac{(N+2)(N-1)}{N}\frac{\varepsilon^2}{7776}+O(\varepsilon^3)\,,
\end{equation}
which matches the free energy in a hypercubic bulk, also computed in \cite{Fei:2015oha}. More interesting than the case when one of the sides of the interface is taken to be free is the RG interface where both CFTs are completely interacting. Setting $\lambda^1_{ijkl}=(4\pi)^2(\delta_{ij}\delta_{kl}+\text{Perms}.)$ and $\lambda^2_{ijkl}=(4\pi)^2\lambda^{B_N}_{ijkl}$, one finds that
\begin{equation}
    \lambda^1_{ijkl}\lambda^2_{ijkl}=\frac{2(N-1)\varepsilon^2}{N+8}(4\pi)^4
\end{equation}
and we obtain the free energy for the $O(N)$--Hypercubic intrface:
\begin{equation}\label{eq:FEOB}
    F_{O(N)-B_N}=\frac{N^4-64N^3-624N^2+896N-128}{N(N+8)^2}\frac{\varepsilon^2}{31104}+O(\varepsilon^3)\,.
\end{equation}
Unlike when one of the sides is free, this interface has a negative free energy for small $N$, until it changes sign at $N\approx 72.44$. We can also examine the two hypertetrahedral CFTs, with interaction tensors taking the form shown in \eqref{eq:lambdaTN} after an appropriate rescaling by $(4\pi)^2$. Using the properties of the $e$-vectors given in \eqref{eq:evecdef}, one can prove the identity
\begin{equation}
    e_i^\alpha e_j^\beta=\delta^{\alpha\beta}-\frac{1}{N+1}\,,
\end{equation}
from which one finds that
\begin{equation}
    \lambda^{T_N}_{ijkl}\lambda^{T_N}_{ijkl}=3N(N+2)\lambda^2+\frac{6N^2}{N+1}\lambda g+\frac{N(N^2-N+1)}{(N+1)^2}g^2\,.
\end{equation}
Plugging this into \eqref{eq:freeenergy}, and using the fixed point values for $\lambda$ and $g$ one finds that the two hypertetrahedral bulks have free energies
\begin{equation}\label{eq:FETet}
    F_{T_N}=\begin{cases}
        -\frac{N(N-1)(N-2)(N^2-6N+11)}{N^2-5N+8}\frac{\varepsilon^2}{7776}+O(\varepsilon^3)\qquad& T_N=T_N^+\\
        -\frac{N(N+1)(N+7)}{(N+3)^2}\frac{\varepsilon^2}{7776}+O(\varepsilon^3)&T_N=T_N^-
    \end{cases}\,.
\end{equation}
Note that $F_{T_N^+}$ is always greater or equal to $F_{T_N^-}$, in accordance with the $F$-theorem, and the instability of the $T_N^+$ fixed point. As there are two inequivalent bulks, one can consider a purely hypertetrahedral RG interface between them, finding a free energy
\begin{equation}
    F_{T_N^+-T_N^-}=\frac{N(-4 N^6+19 N^5-22 N^4-14
   N^3+12 N^2-93 N+358)}{(N+3)^2(N^2-5N+8)^2}\frac{\varepsilon^2}{31104}+O(\varepsilon^3)\,.
\end{equation}
This value is always larger than either of the free energies in \eqref{eq:FETet}, and unlike \eqref{eq:FEOB} is negative for all $N\gtrsim2.504$. When one side of the interface is instead taken to be the $O(N)$ CFT, one finds the free energies
\begin{equation}
    F_{T_N-O(N)}=\begin{cases}
        \frac{N(N^6-20 N^5-265 N^4+2548
   N^3-7624 N^2+8704 N-2048)}{(N+8)^2(N^2-5N+8)^2}\frac{\varepsilon^2}{31104 }+O(\varepsilon^3)\qquad & T_N=T_N^+\\
       \frac{N^4-57 N^3-881 N^2-2589
   N-1658}{(N+8)^2(N+3)^2}\frac{\varepsilon^2}{31104 }+O(\varepsilon^3) & T_N=T_N^-
    \end{cases}\,.
\end{equation}
These free energies once again become positive for sufficiently large $N$. Finally, to determine the free energy for the Hypercubic-Hypertetrahedral interface, we must evaluate the sum
\begin{equation}
    \sum_{i=1}^N\sum_{\alpha=1}^{N+1}(e^\alpha_i)^4\,,
\end{equation}
which appears in the product $\lambda^{B_N}_{ijkl}\lambda^{T_N}_{ijkl}$. Let us first focus on a given term in the sum over $i$, and notice that the definition of the $e$-vectors implies that we can truncate the sum over $\alpha$ at $i+1$. That is,
\begin{equation}
    \sum_{\alpha=1}^{N+1}(e^\alpha_i)^4=\sum_{\alpha=1}^i\left(-\frac{1}{\sqrt{i(i+1)}}\right)^4+\left(\sqrt{\frac{i}{i+1}}\right)^4=1+\frac{1}{i}-\frac{3}{1+i}\,.
\end{equation}
The remaining sum over $i$ is then trivial to perform to find
\begin{equation}
    \sum_{i=1}^N\sum_{\alpha=1}^{N+1}(e^\alpha_i)^4=\frac{N(N+4)}{N+1}-2H_N\,,
\end{equation}
where $H_N$ is the $N^{\text{th}}$ Harmonic number. Then, the product of hypercubic and hypertetrahedral tensors is
\begin{equation}
    \lambda^{B_N}_{ijkl}\lambda^{T_N}_{ijkl}=(2N-2)\varepsilon\lambda+\left(\frac{N^2+3N-16}{3(N+1)}-\frac{2(N-4)}{3N}H_N\right)\varepsilon g\,.
\end{equation}
One then finds
\begin{equation}
    F_{B_N-T_N}=\begin{cases}\left(\frac{(N+1) (N-4)^2
   H_N}{2592 N ((N-5) N+8)}+\frac{N (-2 N^5+9 N^4+57 N^3-512
   N^2+1360 N-1280)-64}{15552
   N ((N-5)
   N+8)^2}\right)\varepsilon^2+O(\varepsilon^3)\qquad &T_N=T_N^+\\
      \left(  \frac{(N-4) (N+1) H_N}{2592
   N (N+3)}-\frac{N (4 N^3+57 N^2+10 N-393)+18}{31104 N
   (N+3)^2}\right)\varepsilon^2+O(\varepsilon^3) & T_N=T_N^-
    \end{cases}\,.
\end{equation}
Both of these free energies are negative for $N\gtrsim2.35$.

\section{Conclusion and Outlook}
\label{sec:discussion}
We have studied localised interactions on an interface between two bulk multiscalar CFTs. At one loop, the defect beta functions in this interface setup can all be obtained from the defect beta function in a single bulk by imagining the bulk to be halfway between the two CFTs. In doing so we have exposed the existence of a considerable number of novel defect universality classes, which, in the previous understanding of the defect beta functions, would have been discarded as mathematical abstractions. Beyond the number of these points, it is also possible to use this construction, by placing two bulks with only a minimal overlap in their global symmetry groups, to produce conformal lines, surfaces and interfaces with considerably less symmetry than those found in previous surveys of fixed points. 

The existence of these new fixed points has implications for the classification of CFTs, and it should be possible to confirm their existence with non-perturbative techniques. Recent advances in bootstrap techniques have now made it possible to effectively isolate line defect CFTs numerically\cite{Lanzetta:2025xfw}. Using these techniques, combined with the previous defect study of \cite{Gliozzi:2015qsa}, in the future one may try to isolate the pinning field defect in the $O(N)$--Free interface, and determine more precisely its distinguishing features relative to the usual pinning field defect. Furthermore, there have been recent development in the bootstrapping of theories with non-maximal symmetry\cite{Kousvos:2025ext}, and one may be able to isolate as well the defects identified in interface systems involving the hypercubic or hypertetrahedral CFTs.

This work has been limited to considering only interfaces between multiscalar CFTs, and then only in $d=4-\varepsilon$. The setup considered here can be immediately generalised to other bulk theories, including those with fermions and with gauge fields, and to theories in higher dimensions. For instance, line defects with fermions in the bulk have been studied before in \cite{Giombi:2022vnz,Pannell:2023pwz}. One can also consider theories with fundamentally different degrees of freedom on either side of the interface, which are not connected by an RG flow. An example of this could be to consider a theory with an exactly marginal bulk operator, and then fixing the two CFTs at different values of this exactly marginal parameter.

Furthermore, one could attempt to understand in more detail properties of the interface theory itself, in the absence of localised interactions. Most notably, there have been a number of recent works detailing the behaviour of conformal defects under fusion\cite{SoderbergRousu:2023zyj,Diatlyk:2024zkk,Diatlyk:2024qpr,Kravchuk:2024qoh}, and it would be interesting to apply those ideas to the case of RG interfaces. The setup with a single RG interface separating two distinct CFTs has a natural generalization in the form of two or more RG interfaces separating a string of CFTs, and it would be interesting to determine whether or not the fusion rules for these defects are as simple as producing the single interface between the two outermost CFTs involved.

In this paper we have also worked out some perturbative conformal data for these RG interfaces, including most notably the interface free energy. Interestingly, we find that the two-point function of $\phi^2$-like operators does not behave as one would expect under the conformal block expansion, once one turns on localised cubic interactions on the boundary. As this conformal block expansion is tied to the folding trick, it would be interesting to understand more precisely how this trick affects degrees of freedom on the interface in the presence of these cubic interactions. It is known that the beta function for boundary theories contains an additional term relative to the beta function for interfaces\cite{Harribey:2023xyv}, associated with a divergence in the tadpole diagram, but a precise correspondence between interface fixed points and boundary fixed points with twice the number of fields of the type implied by the folding trick has not been determined. Finally, it would be interesting to use this CFT data to develop further generalisations of results for RG interfaces in two dimensions, for instance by constructing reflection and transmission coefficients\cite{Quella:2006de} from correlation functions.

\section*{Acknowledgements}
We would like to thank Elizabeth Helfenberger, Christopher Herzog, Edoardo Lauria, and Andreas Stergiou for discussions on the topic. We would like to especially thank Christopher Herzog and Andreas Stergiou for looking at a draft of this paper. SB is supported in part by the STFC under grant ST/X000753/1 and EPSRC under grant EP/Z000580/1.

\begin{appendix}
\section{Details of Bulk Models}
\label{sec:appendix}

The critical bulk models considered in the body of this paper are all fixed points of the one-loop bulk beta function
\begin{equation}\label{eq:betabulk}
    \beta^\lambda_{ijkl}=-\varepsilon\lambda_{ijkl}+(\lambda_{ijmn}\lambda_{mnkl}+\text{Perms.})\,,
\end{equation}
where the additional terms are the two inequivalent permutations of the external indices. Fixed points of this beta function have been studied extensively in \cite{Osborn:2017ucf,Osborn:2020cnf}, but in this paper we focus on three bulk models in particular: the $O(N)$ model, the hypercubic model, and the hypertetrahedral model. For the $O(N)$ model, the fixed point solution to (\ref{eq:betabulk}) is given by
\begin{equation}\label{eq:lambdaON}
\lambda^{O(N)}_{ijkl}=\frac{\varepsilon}{N+8}(\delta_{ij}\delta_{kl}+\text{Perms.}),.
\end{equation}
Note that $O(1)=I$ is the Ising fixed point. The hypercubic fixed point has the symmetries of the hypercube: $B_N=\mathbb{Z}_2^N\rtimes S_N$. To construct $\lambda_{ijkl}$ for this fixed point, we use the two invariant tensors of this group, $\delta_{ij}$ and $\delta_{ijkl}$, and one finds
\begin{equation}\label{eq:lambdaBN}
    \lambda^{B_N}_{ijkl}=\frac{\varepsilon}{3N}(\delta_{ij}\delta_{kl}+\text{Perms}.)+\frac{(N-4)\varepsilon}{3N}\delta_{ijkl}\,.
\end{equation}
For low values of $N$ there are some accidental isometries between this fixed point and other fixed points, $B_1\simeq I$, $B_2\simeq I\times I$, $B_4\simeq O(4)$, where the last isometry holds only at one loop. Finally, the hypertetrahedral fixed point has the symmetries of the hypertetrahedron: $T_N=S_{N+1}\times \mathbb{Z}_2$. It is most natural to write $\lambda_{ijkl}$ at this fixed point in terms of the $N+1$ vectors, $e^\alpha_i$, lying at the vertices of the $N$-Hypertetrahedron\cite{Osborn:2017ucf}. These vectors may be defined using the recursion relations
\begin{equation}\label{eq:evecdef}
\begin{aligned}
    (e_N)_i^\alpha &= (e_{N-1})_i^\alpha\,, &\qquad i&=1,\ldots,N-1,\;\alpha=1,\ldots,N\,,\\
    (e_N)_N^\alpha&=-\sqrt{\frac{1}{N(N+1)}}\,, &\qquad \alpha&=1,\ldots, N\,,\\
    (e_N)_i^{N+1}&=\sqrt{\frac{N}{N+1}}\delta_i{\hspace{-0.4pt}}^N\,, &&{}
\end{aligned}
\end{equation}
with $e^1_1=-\tfrac{1}{\sqrt{2}}=-e^2_1$. Then, $\lambda^{T_N}_{ijkl}$ can be written as
\begin{equation}\label{eq:lambdaTN}
    \lambda^{T_N}_{ijkl}=\lambda(\delta_{ij}\delta_{kl}+\text{Perms.})+g\sum_{\alpha=1}^{N+1}e^\alpha_i e^\alpha_j e^\alpha_k e^\alpha_l\,.
\end{equation}
There are two fixed points, which we will call $T_N^+$ and $T_N^-$ respectively, with this symmetry group, where $\lambda$ and $g$ take the values
\begin{equation}
\begin{aligned}
    \lambda^+&=\frac{\varepsilon}{3(N^2-5N+8)} & \qquad g^+&=\frac{(N-4)(N+1)\varepsilon}{3(N^2-5N+8)}\,,\\ 
    \lambda^-&=\frac{\varepsilon}{3(N+3)} & \;\, g^-&=\frac{(N+1)\varepsilon}{3(N+3)}\,.
\end{aligned}
\end{equation}
There are again accidental isometries at low values of $N$. For $N\leq3$, these fixed points are isomorphic to other fixed points with enhanced symmetry, for $N=4$ $T_4^+\simeq O(4)$, and for $N=5$ $T_5^-\simeq T_5^+$. As $T_N^+$ will be unstable with respect to deformations towards $T_N^-$, we will focus explicitly on the latter theory.
\end{appendix}
\bibliographystyle{JHEP}
\bibliography{Refs} 

\providecommand{\href}[2]{#2}\begingroup\raggedright\begin{thebibliography}{10}

\bibitem{sigl1986order}
L.~Sigl and W.~Fenzl, \emph{Order-parameter exponent $\beta_1$ of a binary liquid mixture at a boundary}, \href{https://doi.org/10.1103/PhysRevLett.57.2191}{\emph{Phys. Rev. Lett.} {\bfseries 57} (1986) 2191}.

\bibitem{mailander1990near}
L.~Mail{\"a}nder, H.~Dosch, J.~Peisl and R.~Johnson, \emph{Near-surface critical x-ray scattering from fe$_\text{3}$al}, \href{https://doi.org/10.1103/PhysRevLett.64.2527}{\emph{Phys. Rev. Lett.} {\bfseries 64} (1990) 2527}.

\bibitem{burandt1993near}
B.~Burandt, W.~Press and S.~Hauss{\"u}hl, \emph{Near-surface x-ray critical scattering from a nh$_\text{4}$br (11$^-$0) surface}, \href{https://doi.org/10.1103/PhysRevLett.71.1188}{\emph{Phys. Rev. Lett.} {\bfseries 71} (1993) 1188}.

\bibitem{alvarado1982surface}
S.~Alvarado, M.~Campagna and H.~Hopster, \emph{Surface magnetism of ni (100) near the critical region by spin-polarized electron scattering}, \href{https://doi.org/10.1103/PhysRevLett.48.51}{\emph{Phys. Rev. Lett.} {\bfseries 48} (1982) 51}.

\bibitem{PhysRevA.19.866}
R.~F. Chang, H.~Burstyn and J.~V. Sengers, \emph{Correlation function near the critical mixing point of a binary liquid}, \href{https://doi.org/10.1103/PhysRevA.19.866}{\emph{Phys. Rev. A} {\bfseries 19} (1979) 866}.

\bibitem{PhysRevB.40.4696}
P.~Damay, F.~Leclercq and P.~Chieux, \emph{Critical scattering function in a binary fluid mixture: A study of sodium-deuteroammonia solution at the critical concentration by small-angle neutron scattering}, \href{https://doi.org/10.1103/PhysRevB.40.4696}{\emph{Phys. Rev. B} {\bfseries 40} (1989) 4696}.

\bibitem{PhysRevB.58.12038}
P.~Damay, F.~Leclercq, R.~Magli, F.~Formisano and P.~Lindner, \emph{Universal critical-scattering function: An experimental approach}, \href{https://doi.org/10.1103/PhysRevB.58.12038}{\emph{Phys. Rev. B} {\bfseries 58} (1998) 12038}.

\bibitem{McAvity:1995zd}
D.~M. McAvity and H.~Osborn, \emph{{Conformal field theories near a boundary in general dimensions}}, \href{https://doi.org/10.1016/0550-3213(95)00476-9}{\emph{Nucl. Phys. B} {\bfseries 455} (1995) 522} [\href{https://arxiv.org/abs/cond-mat/9505127}{{\ttfamily cond-mat/9505127}}].

\bibitem{Billo:2016cpy}
M.~Bill\`o, V.~Gon\c{c}alves, E.~Lauria and M.~Meineri, \emph{{Defects in conformal field theory}}, \href{https://doi.org/10.1007/JHEP04(2016)091}{\emph{JHEP} {\bfseries 04} (2016) 091} [\href{https://arxiv.org/abs/1601.02883}{{\ttfamily 1601.02883}}].

\bibitem{AJBray_1977}
A.~J. Bray and M.~A. Moore, \emph{Critical behaviour of semi-infinite systems}, \href{https://doi.org/10.1088/0305-4470/10/11/021}{\emph{Journal of Physics A: Mathematical and General} {\bfseries 10} (1977) 1927}.

\bibitem{Ohno:1983lma}
K.~Ohno and Y.~Okabe, \emph{{The $1/N$ expansion for the $N$ vector model in the semi-infinite space}}, \href{https://doi.org/10.1143/PTP.70.1226}{\emph{Prog. Theor. Phys.} {\bfseries 70} (1983) 1226}.

\bibitem{gompper1985conformal}
G.~Gompper and H.~Wagner, \emph{Conformal invariance in semi-infinite systems: Application to critical surface scattering}, \href{https://doi.org/10.1007/BF01725537}{\emph{"Zeitschrift f{\"u}r Physik B Condensed Matter"} {\bfseries 59} (1985) 193}.

\bibitem{Diehl:1996kd}
H.~W. Diehl, \emph{{The Theory of Boundary Critical Phenomena}}, \href{https://doi.org/10.1142/S0217979297001751}{\emph{Int. J. Mod. Phys. B} {\bfseries 11} (1997) 3503} [\href{https://arxiv.org/abs/cond-mat/9610143}{{\ttfamily cond-mat/9610143}}].

\bibitem{Metlitski:2020cqy}
M.~A. Metlitski, \emph{{Boundary criticality of the O(N) model in d = 3 critically revisited}}, \href{https://doi.org/10.21468/SciPostPhys.12.4.131}{\emph{SciPost Phys.} {\bfseries 12} (2022) 131} [\href{https://arxiv.org/abs/2009.05119}{{\ttfamily 2009.05119}}].

\bibitem{Padayasi:2021sik}
J.~Padayasi, A.~Krishnan, M.~A. Metlitski, I.~A. Gruzberg and M.~Meineri, \emph{{The extraordinary boundary transition in the 3d O(N) model via conformal bootstrap}}, \href{https://doi.org/10.21468/SciPostPhys.12.6.190}{\emph{SciPost Phys.} {\bfseries 12} (2022) 190} [\href{https://arxiv.org/abs/2111.03071}{{\ttfamily 2111.03071}}].

\bibitem{Toldin:2021kun}
F.~P. Toldin and M.~A. Metlitski, \emph{{Boundary Criticality of the 3D O(N) Model: From Normal to Extraordinary}}, \href{https://doi.org/10.1103/PhysRevLett.128.215701}{\emph{Phys. Rev. Lett.} {\bfseries 128} (2022) 215701} [\href{https://arxiv.org/abs/2111.03613}{{\ttfamily 2111.03613}}].

\bibitem{Krishnan:2023cff}
A.~Krishnan and M.~A. Metlitski, \emph{{A plane defect in the 3d O(N) model}}, \href{https://doi.org/10.21468/SciPostPhys.15.3.090}{\emph{SciPost Phys.} {\bfseries 15} (2023) 090} [\href{https://arxiv.org/abs/2301.05728}{{\ttfamily 2301.05728}}].

\bibitem{DeSabbata:2025ano}
E.~De~Sabbata, \emph{{Defects in Conformal Field Theory with O(N) Symmetry}}, Ph.D. thesis, Universit{\`a} degli Studi di Torino, Universit{\`a} degli Studi di Torino, Italy, 2025.

\bibitem{Giombi:2023dqs}
S.~Giombi and B.~Liu, \emph{{Notes on a surface defect in the O(N) model}}, \href{https://doi.org/10.1007/JHEP12(2023)004}{\emph{JHEP} {\bfseries 12} (2023) 004} [\href{https://arxiv.org/abs/2305.11402}{{\ttfamily 2305.11402}}].

\bibitem{Trepanier:2023tvb}
M.~Tr\'epanier, \emph{{Surface defects in the O(N) model}}, \href{https://doi.org/10.1007/JHEP09(2023)074}{\emph{JHEP} {\bfseries 09} (2023) 074} [\href{https://arxiv.org/abs/2305.10486}{{\ttfamily 2305.10486}}].

\bibitem{Raviv-Moshe:2023yvq}
A.~Raviv-Moshe and S.~Zhong, \emph{{Phases of surface defects in Scalar Field Theories}}, \href{https://doi.org/10.1007/JHEP08(2023)143}{\emph{JHEP} {\bfseries 08} (2023) 143} [\href{https://arxiv.org/abs/2305.11370}{{\ttfamily 2305.11370}}].

\bibitem{Harribey:2023xyv}
S.~Harribey, I.~R. Klebanov and Z.~Sun, \emph{{Boundaries and interfaces with localized cubic interactions in the O(N) model}}, \href{https://doi.org/10.1007/JHEP10(2023)017}{\emph{JHEP} {\bfseries 10} (2023) 017} [\href{https://arxiv.org/abs/2307.00072}{{\ttfamily 2307.00072}}].

\bibitem{deSabbata:2024xwn}
E.~de~Sabbata, N.~Drukker and A.~Stergiou, \emph{{Transdimensional Defects}},  \href{https://arxiv.org/abs/2411.17809}{{\ttfamily 2411.17809}}.

\bibitem{Bianchi:2024eqm}
L.~Bianchi, L.~S. Cardinale and E.~de~Sabbata, \emph{{Defects in the long-range O(N) model}},  \href{https://arxiv.org/abs/2412.08697}{{\ttfamily 2412.08697}}.

\bibitem{Pelissetto:2000ek}
A.~Pelissetto and E.~Vicari, \emph{{Critical phenomena and renormalization group theory}}, \href{https://doi.org/10.1016/S0370-1573(02)00219-3}{\emph{Phys. Rept.} {\bfseries 368} (2002) 549} [\href{https://arxiv.org/abs/cond-mat/0012164}{{\ttfamily cond-mat/0012164}}].

\bibitem{Osborn:2017ucf}
H.~Osborn and A.~Stergiou, \emph{{Seeking fixed points in multiple coupling scalar theories in the $\varepsilon$ expansion}}, \href{https://doi.org/10.1007/JHEP05(2018)051}{\emph{JHEP} {\bfseries 05} (2018) 051} [\href{https://arxiv.org/abs/1707.06165}{{\ttfamily 1707.06165}}].

\bibitem{Rychkov:2018vya}
S.~Rychkov and A.~Stergiou, \emph{General properties of multiscalar {RG} flows in $d=4-\varepsilon$}, \href{https://doi.org/10.21468/SciPostPhys.6.1.008}{\emph{SciPost Phys.} {\bfseries 6} (2019) 008} [\href{https://arxiv.org/abs/1810.10541}{{\ttfamily 1810.10541}}].

\bibitem{Osborn:2020cnf}
H.~Osborn and A.~Stergiou, \emph{{Heavy handed quest for fixed points in multiple coupling scalar theories in the $\varepsilon$ expansion}}, \href{https://doi.org/10.1007/JHEP04(2021)128}{\emph{JHEP} {\bfseries 04} (2021) 128} [\href{https://arxiv.org/abs/2010.15915}{{\ttfamily 2010.15915}}].

\bibitem{Pannell:2023pwz}
W.~H. Pannell and A.~Stergiou, \emph{{Line defect RG flows in the \ensuremath{\varepsilon} expansion}}, \href{https://doi.org/10.1007/JHEP06(2023)186}{\emph{JHEP} {\bfseries 06} (2023) 186} [\href{https://arxiv.org/abs/2302.14069}{{\ttfamily 2302.14069}}].

\bibitem{Anataichuk:2025zoq}
A.~Anataichuk and S.~Harribey, \emph{{Note on surface defects in multiscalar critical models}}, \href{https://doi.org/10.1088/1751-8121/adf26d}{\emph{J. Phys. A} {\bfseries 58} (2025) 315403} [\href{https://arxiv.org/abs/2503.05519}{{\ttfamily 2503.05519}}].

\bibitem{Harribey:2024gjn}
S.~Harribey, W.~H. Pannell and A.~Stergiou, \emph{{Multiscalar Critical Models with Localised Cubic Interactions}},  \href{https://arxiv.org/abs/2407.20326}{{\ttfamily 2407.20326}}.

\bibitem{Pannell:2024hbu}
W.~H. Pannell, \emph{{A note on defect stability in d = 4 {\ensuremath{-}} {\ensuremath{\varepsilon}}}}, \href{https://doi.org/10.1007/JHEP12(2024)187}{\emph{JHEP} {\bfseries 12} (2024) 187} [\href{https://arxiv.org/abs/2408.15315}{{\ttfamily 2408.15315}}].

\bibitem{Gliozzi:2015qsa}
F.~Gliozzi, P.~Liendo, M.~Meineri and A.~Rago, \emph{{Boundary and Interface CFTs from the Conformal Bootstrap}}, \href{https://doi.org/10.1007/JHEP05(2015)036}{\emph{JHEP} {\bfseries 05} (2015) 036} [\href{https://arxiv.org/abs/1502.07217}{{\ttfamily 1502.07217}}].

\bibitem{Giombi:2024qbm}
S.~Giombi, E.~Helfenberger and H.~Khanchandani, \emph{{RG Interfaces from Double-Trace Deformations}},  \href{https://arxiv.org/abs/2407.07856}{{\ttfamily 2407.07856}}.

\bibitem{Gaiotto:2012np}
D.~Gaiotto, \emph{{Domain Walls for Two-Dimensional Renormalization Group Flows}}, \href{https://doi.org/10.1007/JHEP12(2012)103}{\emph{JHEP} {\bfseries 12} (2012) 103} [\href{https://arxiv.org/abs/1201.0767}{{\ttfamily 1201.0767}}].

\bibitem{Quella:2006de}
T.~Quella, I.~Runkel and G.~M.~T. Watts, \emph{{Reflection and transmission for conformal defects}}, \href{https://doi.org/10.1088/1126-6708/2007/04/095}{\emph{JHEP} {\bfseries 04} (2007) 095} [\href{https://arxiv.org/abs/hep-th/0611296}{{\ttfamily hep-th/0611296}}].

\bibitem{Cogburn:2023xzw}
C.~V. Cogburn, A.~L. Fitzpatrick and H.~Geng, \emph{{CFT and lattice correlators near an RG domain wall between minimal models}}, \href{https://doi.org/10.21468/SciPostPhysCore.7.2.021}{\emph{SciPost Phys. Core} {\bfseries 7} (2024) 021} [\href{https://arxiv.org/abs/2308.00737}{{\ttfamily 2308.00737}}].

\bibitem{Konechny:2020jym}
A.~Konechny, \emph{{Properties of RG interfaces for 2D boundary flows}}, \href{https://doi.org/10.1007/JHEP05(2021)178}{\emph{JHEP} {\bfseries 05} (2021) 178} [\href{https://arxiv.org/abs/2012.12361}{{\ttfamily 2012.12361}}].

\bibitem{HultgreenMena:2025uok}
C.~W. Hultgreen~Mena, \emph{{Janus and RG-flow interfaces in gauged supergravity}}, Ph.D. thesis, UCLA, Los Angeles (main), 2025.

\bibitem{Lanzetta:2025xfw}
R.~A. Lanzetta, S.~Liu and M.~A. Metlitski, \emph{{The beginning of the endpoint bootstrap for conformal line defects}},  \href{https://arxiv.org/abs/2508.14964}{{\ttfamily 2508.14964}}.

\bibitem{Myers:2010xs}
R.~C. Myers and A.~Sinha, \emph{{Seeing a c-theorem with holography}}, \href{https://doi.org/10.1103/PhysRevD.82.046006}{\emph{Phys. Rev. D} {\bfseries 82} (2010) 046006} [\href{https://arxiv.org/abs/1006.1263}{{\ttfamily 1006.1263}}].

\bibitem{Jafferis:2011zi}
D.~L. Jafferis, I.~R. Klebanov, S.~S. Pufu and B.~R. Safdi, \emph{{Towards the F-Theorem: N=2 Field Theories on the Three-Sphere}}, \href{https://doi.org/10.1007/JHEP06(2011)102}{\emph{JHEP} {\bfseries 06} (2011) 102} [\href{https://arxiv.org/abs/1103.1181}{{\ttfamily 1103.1181}}].

\bibitem{Klebanov:2011gs}
I.~R. Klebanov, S.~S. Pufu and B.~R. Safdi, \emph{{F-Theorem without Supersymmetry}}, \href{https://doi.org/10.1007/JHEP10(2011)038}{\emph{JHEP} {\bfseries 10} (2011) 038} [\href{https://arxiv.org/abs/1105.4598}{{\ttfamily 1105.4598}}].

\bibitem{Casini:2012ei}
H.~Casini and M.~Huerta, \emph{{On the RG running of the entanglement entropy of a circle}}, \href{https://doi.org/10.1103/PhysRevD.85.125016}{\emph{Phys. Rev. D} {\bfseries 85} (2012) 125016} [\href{https://arxiv.org/abs/1202.5650}{{\ttfamily 1202.5650}}].

\bibitem{Fei:2015oha}
L.~Fei, S.~Giombi, I.~R. Klebanov and G.~Tarnopolsky, \emph{{Generalized $F$-Theorem and the $\epsilon$ Expansion}}, \href{https://doi.org/10.1007/JHEP12(2015)155}{\emph{JHEP} {\bfseries 12} (2015) 155} [\href{https://arxiv.org/abs/1507.01960}{{\ttfamily 1507.01960}}].

\bibitem{Pannell:2025ixz}
W.~H. Pannell and A.~Stergiou, \emph{{Gradient flows and the curvature of theory space}}, \href{https://doi.org/10.1007/JHEP09(2025)117}{\emph{JHEP} {\bfseries 09} (2025) 117} [\href{https://arxiv.org/abs/2502.06940}{{\ttfamily 2502.06940}}].

\bibitem{Affleck:1991tk}
I.~Affleck and A.~W.~W. Ludwig, \emph{{Universal noninteger 'ground state degeneracy' in critical quantum systems}}, \href{https://doi.org/10.1103/PhysRevLett.67.161}{\emph{Phys. Rev. Lett.} {\bfseries 67} (1991) 161}.

\bibitem{Friedan:2003yc}
D.~Friedan and A.~Konechny, \emph{{On the boundary entropy of one-dimensional quantum systems at low temperature}}, \href{https://doi.org/10.1103/PhysRevLett.93.030402}{\emph{Phys. Rev. Lett.} {\bfseries 93} (2004) 030402} [\href{https://arxiv.org/abs/hep-th/0312197}{{\ttfamily hep-th/0312197}}].

\bibitem{Casini:2016fgb}
H.~Casini, I.~Salazar~Landea and G.~Torroba, \emph{{The g-theorem and quantum information theory}}, \href{https://doi.org/10.1007/JHEP10(2016)140}{\emph{JHEP} {\bfseries 10} (2016) 140} [\href{https://arxiv.org/abs/1607.00390}{{\ttfamily 1607.00390}}].

\bibitem{Yamaguchi:2002pa}
S.~Yamaguchi, \emph{{Holographic RG flow on the defect and g theorem}}, \href{https://doi.org/10.1088/1126-6708/2002/10/002}{\emph{JHEP} {\bfseries 10} (2002) 002} [\href{https://arxiv.org/abs/hep-th/0207171}{{\ttfamily hep-th/0207171}}].

\bibitem{Fujita:2011fp}
M.~Fujita, T.~Takayanagi and E.~Tonni, \emph{{Aspects of AdS/BCFT}}, \href{https://doi.org/10.1007/JHEP11(2011)043}{\emph{JHEP} {\bfseries 11} (2011) 043} [\href{https://arxiv.org/abs/1108.5152}{{\ttfamily 1108.5152}}].

\bibitem{Nozaki:2012qd}
M.~Nozaki, T.~Takayanagi and T.~Ugajin, \emph{{Central Charges for BCFTs and Holography}}, \href{https://doi.org/10.1007/JHEP06(2012)066}{\emph{JHEP} {\bfseries 06} (2012) 066} [\href{https://arxiv.org/abs/1205.1573}{{\ttfamily 1205.1573}}].

\bibitem{Gaiotto:2014gha}
D.~Gaiotto, \emph{{Boundary F-maximization}},  \href{https://arxiv.org/abs/1403.8052}{{\ttfamily 1403.8052}}.

\bibitem{Estes:2014hka}
J.~Estes, K.~Jensen, A.~O'Bannon, E.~Tsatis and T.~Wrase, \emph{{On Holographic Defect Entropy}}, \href{https://doi.org/10.1007/JHEP05(2014)084}{\emph{JHEP} {\bfseries 05} (2014) 084} [\href{https://arxiv.org/abs/1403.6475}{{\ttfamily 1403.6475}}].

\bibitem{Jensen:2015swa}
K.~Jensen and A.~O'Bannon, \emph{{Constraint on Defect and Boundary Renormalization Group Flows}}, \href{https://doi.org/10.1103/PhysRevLett.116.091601}{\emph{Phys. Rev. Lett.} {\bfseries 116} (2016) 091601} [\href{https://arxiv.org/abs/1509.02160}{{\ttfamily 1509.02160}}].

\bibitem{Herzog:2017kkj}
C.~Herzog, K.-W. Huang and K.~Jensen, \emph{{Displacement Operators and Constraints on Boundary Central Charges}}, \href{https://doi.org/10.1103/PhysRevLett.120.021601}{\emph{Phys. Rev. Lett.} {\bfseries 120} (2018) 021601} [\href{https://arxiv.org/abs/1709.07431}{{\ttfamily 1709.07431}}].

\bibitem{Kobayashi:2018lil}
N.~Kobayashi, T.~Nishioka, Y.~Sato and K.~Watanabe, \emph{{Towards a $C$-theorem in defect CFT}}, \href{https://doi.org/10.1007/JHEP01(2019)039}{\emph{JHEP} {\bfseries 01} (2019) 039} [\href{https://arxiv.org/abs/1810.06995}{{\ttfamily 1810.06995}}].

\bibitem{Casini:2018nym}
H.~Casini, I.~Salazar~Landea and G.~Torroba, \emph{{Irreversibility in quantum field theories with boundaries}}, \href{https://doi.org/10.1007/JHEP04(2019)166}{\emph{JHEP} {\bfseries 04} (2019) 166} [\href{https://arxiv.org/abs/1812.08183}{{\ttfamily 1812.08183}}].

\bibitem{Giombi:2020rmc}
S.~Giombi and H.~Khanchandani, \emph{{CFT in AdS and boundary RG flows}}, \href{https://doi.org/10.1007/JHEP11(2020)118}{\emph{JHEP} {\bfseries 11} (2020) 118} [\href{https://arxiv.org/abs/2007.04955}{{\ttfamily 2007.04955}}].

\bibitem{Kousvos:2025ext}
S.~R. Kousvos and A.~Stergiou, \emph{{Redundancy Channels in the Conformal Bootstrap}},  \href{https://arxiv.org/abs/2507.05338}{{\ttfamily 2507.05338}}.

\bibitem{Giombi:2022vnz}
S.~Giombi, E.~Helfenberger and H.~Khanchandani, \emph{{Line defects in fermionic CFTs}}, \href{https://doi.org/10.1007/JHEP08(2023)224}{\emph{JHEP} {\bfseries 08} (2023) 224} [\href{https://arxiv.org/abs/2211.11073}{{\ttfamily 2211.11073}}].

\bibitem{SoderbergRousu:2023zyj}
A.~S{\"o}derberg~Rousu, \emph{{Fusion of conformal defects in interacting theories}}, \href{https://doi.org/10.1007/JHEP10(2023)183}{\emph{JHEP} {\bfseries 10} (2023) 183} [\href{https://arxiv.org/abs/2304.10239}{{\ttfamily 2304.10239}}].

\bibitem{Diatlyk:2024zkk}
O.~Diatlyk, H.~Khanchandani, F.~K. Popov and Y.~Wang, \emph{{Defect fusion and Casimir energy in higher dimensions}}, \href{https://doi.org/10.1007/JHEP09(2024)006}{\emph{JHEP} {\bfseries 09} (2024) 006} [\href{https://arxiv.org/abs/2404.05815}{{\ttfamily 2404.05815}}].

\bibitem{Diatlyk:2024qpr}
O.~Diatlyk, H.~Khanchandani, F.~K. Popov and Y.~Wang, \emph{{Effective Field Theory of Conformal Boundaries}},  \href{https://arxiv.org/abs/2406.01550}{{\ttfamily 2406.01550}}.

\bibitem{Kravchuk:2024qoh}
P.~Kravchuk, A.~Radcliffe and R.~Sinha, \emph{{Effective theory for fusion of conformal defects}},  \href{https://arxiv.org/abs/2406.04561}{{\ttfamily 2406.04561}}.

\end{thebibliography}\endgroup

\addcontentsline{toc}{section}{References}

\end{document}